\documentclass[11pt]{article}
\usepackage{amsmath,amssymb,amsbsy,amsthm}

\let\epsilon\varepsilon
\let\phi\varphi

\newtheorem{theorem}{Theorem}
\newtheorem{lemma}{Lemma}

\newtheorem{claim}{Claim}

\begin{document}
\title{Applications of Universal Source Coding to
Statistical Analysis of Time Series }
\author{\large Boris Ryabko \\  \\
Siberian State University of Telecommunications and
Informatics\\
and Institute of Computational Technology of Siberian Branch of
\\ Russian Academy of Science \\
  boris@ryabko.net
   }
\date{}
\maketitle

\begin{abstract}
We show how
 universal codes   can be
used for solving some of the most important statistical problems
for time series. By definition,
 a universal code (or a universal lossless data compressor) can compress
any sequence generated by a stationary and ergodic source
asymptotically to the Shannon entropy, which, in turn, is the best
achievable   ratio for  lossless data compressors.

 We
consider finite-alphabet and real-valued time series
 and  the  following  problems:
   estimation of the limiting probabilities for finite-alphabet time series and  estimation of the
density  for real-valued time series,  the  on-line prediction,
regression,   classification (or  problems with side information)
for both types of the time series and the following  problems of
hypothesis testing: goodness-of-fit testing, or identity testing,
and testing of serial independence. It is important to note that
all problems
 are considered in the
 framework of classical mathematical
statistics and, on the other hand, everyday methods of data
compression (or archivers) can be used as a tool for the
estimation and testing.

It turns out, that quite often the suggested methods and tests are
more powerful than known ones when they are applied in practice.

\end{abstract}




\section{Introduction}\label{sec1.1}
Since C. Shannon  published the paper ``A mathematical theory of
communication''  \cite{S}, the ideas and results of Information
Theory have played 
an important role in cryptography \cite{Ma, S2}, mathematical
statistics \cite{Ba, CS, Ku},
and many other fields \cite{V1,V2}, which are far from
telecommunications. Universal coding, which is a part of
Information Theory, also has been efficiently applied in
 many fields since its discovery \cite{K, Fi}. Thus, application of
results of universal coding, initiated in 1988 $\,$ \cite{ Ry1},
created a new approach to prediction \cite{Al,Ki, MM, No}. Maybe
the most unexpected application of data compression ideas arises
in experiments that show that some ant species are capable of 
compressing messages and are capable of adding and subtracting
small numbers \cite{Rez,RR}.

In this chapter we describe a new approach to estimation,
prediction and hypothesis testing for time series, which was
suggested recently  \cite{Ry1,RA,RM}. This approach is based on
ideas of universal coding (or universal data compression). We
would like to emphasize that  everyday methods of data compression
(or archivers) can be directly used as a tool for estimation and
hypothesis testing. It is important to note that the modern
archivers (like \emph{zip}, \emph{arj, rar,} etc.) are based on
deep theoretical results of the source coding theory \cite{e,
K-Y,Kr,Ri,Sa} and have shown their high efficiency in practice
because
 archivers can find many kinds of latent regularities and use them for compression.

 It is worth noting
that this approach was applied to the problem
 of randomness testing \cite{RM}. This problem is quite important for practice;
  in particular,  the National Institute of Standards and Technology
   of  USA (NIST)
 has suggested ``A statistical test suite for random
and pseudorandom number generators for cryptographic
applications'' \cite{rng}, which consists of 16 tests. It has
turned
 out that tests
 which are based on universal codes are more powerful than the tests suggested by  NIST
$ \,$   \cite{RM}.

The outline of this  paper
  is as follows. The next section contains
some necessary definitions and facts about predictors,  codes,
hypothesis testing and description of one universal  code. The
section~\ref{s:21} and~\ref{s:22}
 are devoted to problems of estimation and
hypothesis testing, correspondingly, for the case of
finite-alphabet time series. The case of infinite alphabets  is
considered in the
~\ref{s:3} section. 
 All proofs are given in Appendix, but some
intuitive indication are given in the body of the paper.

\section{Definitions and Statements of the Problems}\label{s:222}

\subsection{Estimation and Prediction for I.I.D. Sources}

First we consider a source with unknown statistics which generates
sequences $x_1x_2\cdots$ of letters from some set (or alphabet)
$A$. It will be convenient  now to  describe briefly the
prediction problem. Let the source generate a message $x_1\ldots
x_{t-1}x_t$, $\, x_i\in A$ for all $i$, and the following letter
$x_{t+1}$ needs to be predicted.
 This problem can be traced back to Laplace  \cite{FE,orl} who considered
the problem of estimation of the
 probability that the sun will
rise tomorrow, given that it has risen every day since Creation.
In our notation the alphabet $A$ contains two letters $ 0 \; ("the
\:sun \:rises") $  and $1 \;("the \:sun\: does\: not\: rise" ) ,$
$t$ is the number of days since Creation, $x_1\ldots x_{t-1}x_t =
00 \ldots 0 .$

Laplace suggested the following predictor:
\begin{equation}\label{L}
L_0(a|x_1\cdots x_t) = (\nu_{x_1\cdots x_t}(a) +1 )/ (t+ |A | ),
\end{equation}   where $\nu_{x_1\cdots x_t}(a)$
denotes the count of letter $a$ occurring in the word $x_1\ldots
x_{t-1}x_t.$ It is important to note that the predicted
probabilities cannot be equal to zero even through a certain
letter did not occur in the word $x_1\ldots x_{t-1}x_t.$

\textbf{Example.} Let $ A= \{0, 1 \}, \: x_1 ... x_5 = 01010,$
then the Laplace prediction is as follows: $L_0(x_{6}=0| x_1 ...
x_5 =01010) = (3+1)/ (5+2) = 4/7, L_0(x_{6}=1 |x_1 ... x_5 =
01010) = (2+1)/ (5+2) = 3/7.$ In other words, $3/7$ and  $4/7 $
are estimations of the unknown probabilities $P(x_{t+1} = 0|x_1
\ldots x_t = 01010 )$ and $P(x_{t+1} = 1 |x_1 \ldots x_t = 01010 )
.$ (In what follows we will use the shorter notation: $P( 0| 01010
)$ and $P( 1 | 01010 ) ).$

We can see that Laplace  considered  prediction as a set of
estimations of unknown (conditional) probabilities. This approach
to the problem of prediction was developed in 1988  \cite{Ry1} and
now is
 often called on-line prediction or universal prediction
 \cite{Al,Ki,MM,No}.
 As we mentioned above, it seems
natural to consider conditional probabilities to be
 the  best prediction, because they contain all information about
 the
 future  behavior of the stochastic process. Moreover, this approach
 is deeply connected with game-theoretical interpretation of
 prediction   \cite{Ke,R94} and, in fact,
 all obtained results can be easily transferred from one model to the other.

               Any predictor $\gamma$  defines a measure (or an estimation of
 probability)  by the following equation
\begin{equation}\label{p} \gamma (x_1 ... x_t) = \prod_{i=1}^t \gamma(x_i | x_1 ... x_{i-1}
).
\end{equation}
 And, vice versa, any measure $
\gamma$ (or estimation of
 probability) defines a predictor:
\begin{equation}\label{p1}\gamma(x_i | x_1 ...\, x_{i-1}) =
 \gamma( x_1 ...\, x_{i-1} x_i) / \gamma( x_1 ...\, x_{i-1} ). \end{equation}
\textbf{Example.} Let us apply the Laplace predictor for
estimation of probabilities  of the sequences  $01010$ and
$010101$. From (\ref{p}) we obtain $L_0(01010) = \frac{1}{2}
\frac{1}{3} \frac{2}{4} \frac{2}{5} \frac{3}{6} = \frac{1}{60},$
$L_0(010101) = \frac{1}{60}\frac{3}{7} = \frac{1}{140}.$ Vice
versa, if for some measure (or a probability  estimation) $\chi$
we have $\chi(01010) =\frac{1}{60}$ and $\chi(010101)
=\frac{1}{140},$ then we obtain from (\ref{p1}) the following
prediction, or the estimation of the conditional probability, $ \:
\chi(1|01010) =\frac{1/140}{1/60}$ $= \frac{3}{7}.$

Now we concretize the class of stochastic processes 
 which will be
considered. Generally speaking, we will deal with so-called
stationary and ergodic time series  (or sources), whose definition
will be given later, but now we  consider may be the simplest
class of such processes, which are called i.i.d. sources.
 By definition, they  generate independent
and identically distributed random variables from some set $A.$ In
our case  $A$ will be either some alphabet or a real-valued
interval.

The next natural question is how to measure the errors   of
prediction and   estimation of probability. Mainly we will measure
these errors by the Kullback-Leibler (KL) divergence which is
defined by
\begin{equation}\label{kl}
D(P,Q) = \sum_{a\in A}P(a) \log \frac{P(a)}{Q(a)} \: ,
\end{equation} where $P(a)$ and $Q(a)$ are probability
distributions over an alphabet $A$ (here and below $\log \equiv
\log_2$
 and $0 \log 0 =
0$). The probability distribution $P(a)$ can be considered as
unknown whereas $Q(a)$ is its estimation. It is well-known that
for any distributions $P$ and $Q$ the KL divergence is nonnegative
and equals $0$ if and only if $P(a) = Q(a)$ for all $a \:$
\cite{Ga}. So, if the estimation $Q$ is equal to $P,$ the error is
$0$, otherwise the error is  a positive number.

 The KL divergence is connected with the
so-called variation distance
$$ || P - Q || = \sum_{a \in A} | P(a) - Q(a)| , $$ via
 the
the following inequality (Pinsker's inequality)
\begin{equation}\label{pi}
\sum_{a\in A} P(a)\log\frac{P(a)} {Q(a)} \geq  \frac{\log e} {2}
|| P - Q ||^2.
\end{equation}
Let $\gamma$ be a predictor, i.e.  an  estimation of an unknown
conditional  probability and $x_1\cdots x_t$ be a sequence of
letters created by an unknown source $P$. The KL divergence
between $P$ and the predictor $\gamma$ is equal  to
\begin{equation}\label{r0}
\rho_{\gamma,P}(x_1\cdots x_t)= \sum_{a\in A}P(a|x_1\cdots
x_t)\log\frac{P(a|x_1\cdots x_t)} {\gamma(a|x_1\cdots x_t)},
\end{equation}
For fixed $t$
 it is a random variable, because $x_1, x_2, \cdots,
x_t$ are random variables. We define the average error
  at time $t$ by
\begin{equation}\label{r1}
\rho^t(P\|\gamma)=E\,\left(\rho_{\gamma,P}(\cdot)\right)= \,
\sum_{x_1\cdots x_t\in A^t}P(x_1\cdots
x_t)\,\,\rho_{\gamma,P}(x_1\cdots x_t) \end{equation}
$$=\sum_{x_1\cdots x_t\in
A^t}P(x_1\cdots x_t)\,\,\sum_{a\in A}P(a|x_1\cdots
x_t)\log\frac{P(a|x_1\cdots x_t)} {\gamma(a|x_1\cdots x_t)}.$$
Analogously,
 if $\gamma(\,)$ is an estimation of a
 probability
distribution we define the errors \emph{per letter} as follows:
\begin{equation}\label{R0} \bar{\rho}_{\gamma, P}(x_1 ... x_t) =
t^{-1}\: ( \log ( P(x_1 ... x_t) / \gamma (x_1 ... x_t) )
\end{equation} and
\begin{equation}\label{RR} \bar{\rho}^t (P\|\gamma) = t^{-1} \sum_{x_1 ... x_t \in A^t} P(x_1 ... x_t)
\log ( P(x_1 ... x_t) / \gamma (x_1 ... x_t) ),
\end{equation} where, as before,
$
 \gamma (x_1 ... x_t) = \prod_{i=1}^t \gamma(x_i | x_1 ... x_{i-1}
). $ (Here and below we denote by $A^t$ and $A^*$ the set of all
words of length $t$ over $A$ and the set of all finite words over
$A$ correspondingly: $A^* = \bigcup_{i=1}^\infty A^i$.)

\begin{claim}[\cite{Ry1}]\label{Lap}
For any i.i.d. source $P$ generating letters from an alphabet $A$
and an integer $t$ the average error (\ref{r1}) of the Laplace
predictor and the average error of the Laplace estimator are upper
bounded as follows:
\begin{equation}\label{rL} \rho^t(P\|L_0)
\leq ( (|A| - 1) \log e ) / (t + 1), \end{equation}
\begin{equation}\label{RL} \bar{\rho}^t(P\|L_0)
\leq (|A| - 1) \log t / t +O(1/t), \end{equation} where $e \simeq
2.718$ is the Euler number.
 \end{claim}
 So, we can see that the average error of the Laplace predictor
 goes to zero for any i.i.d. source $P$ when the length $t$ of the sample $x_1\cdots x_t$
 tends to infinity.
Such methods  are called universal, because the error goes to zero
for any source, or process. In this case they are universal for
the set of all i.i.d. sources generating letters from the finite
alphabet $A,$ but later we consider universal estimators for the
set of stationary and ergodic sources. It is worth noting  that
the first universal code for which the estimation (\ref{RL}) is
valid, was suggested independently by Fitingof \cite{Fi} and
Kolmogorov \cite{K} in 1966.

The value
$$\bar{\rho}^t(P\|\gamma) =  t^{-1} \sum_{x_1 ... x_t \in A^t} P(x_1 ...
x_t) \log ( P(x_1 ... x_t) / \gamma (x_1 ... x_t) )$$ has one more
interpretation connected with data compression. Now we consider
the main idea whereas the more formal definitions will be given
later.  First we recall the definition of the Shannon entropy
$h_0(P)$ for an i.i.d. source $P$
\begin{equation}\label{h0} h_0(P) = -  \sum_{a \in A} P(a) \log P(a). \end{equation}
It is easy to see that $ t^{-1} \sum_{x_1 ... x_t \in A^t} P(x_1
... x_t) \log ( P(x_1 ... x_t)) $ $= - h_0(P)$ for the i.i.d.
source.
Hence, we can represent 
 the
average error $\bar{\rho}^t(P\|\gamma)$ in (\ref{RR}) as
$$\bar{\rho}^t(P\|\gamma) =  t^{-1} \sum_{x_1 ... x_t \in A^t} P(x_1 ...
x_t) \log ( 1 / \gamma (x_1 ... x_t) ) - h_0(P).$$ More formal and
general consideration of universal codes will be given later, but
here we briefly show how estimations and codes are connected. The
point is that one can construct a code with codelength
$\gamma_{code}(a|x_1\cdots x_t)\approx -\log_2 \gamma(a|x_1\cdots
x_n)$ for any letter $a\in A$ (since Shannon's original research,
it has been well known that, using block codes with large block
length or more modern methods of arithmetic coding \cite{Ri1} ,
the approximation may be as accurate as you
like). If one knows the real distribution $P,$ one 
can base coding on the true distribution $P$ and not on the
prediction $\gamma$. The difference in performance measured by
average code length is given by
$$
\sum_{a\in A}P(a|x_1\cdots x_t)(-\log_2 \gamma (a|x_1\cdots x_t))
-\sum_{a\in A}P(a|x_1\cdots x_t)(-\log_2 P(a|x_1\cdots x_t))$$
$$
=\sum_{a\in A}P(a|x_1\cdots x_t)\log_2 \frac{P(a|x_1\cdots x_t)}
{\gamma (a|x_1\cdots x_t)}.$$
 Thus this excess  is exactly the
error defined above (\ref{r0}) . Analogously, if we encode the
sequence $x_1 \ldots x_t $ based on a predictor $\gamma$ the
redundancy per letter
 is defined by (\ref{R0})
and (\ref{RR}). So, from mathematical point of view, the
estimation of the limiting probabilities  and universal coding are
identical. But $- \, \log \gamma (x_1 ... x_t) $ and $- \,\log P
(x_1 ... x_t) $ have a very natural interpretation. The first
value is a code word length (in bits), if the "code" $\gamma$ is
applied for compressing  the
word $x_1 ... x_t$ and the second one is the minimal possible 
codeword length. The difference is the redundancy of the code and,
at the same time, the error of the predictor. It is worth noting
that there are many other deep interrelations between the
universal coding, prediction and estimation  \cite{Ri,Ry1}.

 We can see from the claim and the
 Pinsker inequality (\ref{pi}) that the variation distance of the Laplace predictor and estimator goes to
 zero, too.
 Moreover, it can be easily shown
 that the error (\ref{r0}) (and the corresponding variation
 distance) goes to zero with probability 1, when $t $ goes to
 infinity. (Informally, it means that the error (\ref{r0}) goes to
 zero for almost all sequences $x_1\cdots x_t$ according to the
 measure $P.$ )
 Obviously, such  properties are very
 desirable for any predictor and for larger classes of sources, like Markov and
  stationary  ergodic (they will be briefly defined in the next subsection).  However, it is proven
  \cite{Ry1}   that such predictors do not exist
for the class of
  all stationary and ergodic sources
  (generating letters from a given finite alphabet).
  More precisely, if, for example, the alphabet has two letters, then 
    for any predictor $\gamma$ and for any $\delta > 0$ there exists a source $P$  such that
   with probability 1 $\:\rho_{\gamma,P}(x_1\cdots
  x_t) \geq \, 1/2 - \delta$ infinitely often when
  $t\rightarrow \infty.$
  In other words,  the error of any predictor may
not go to 0, if the predictor is applied to an arbitrary
stationary and ergodic source,
 that is why it is difficult to use (\ref{r0}) and
(\ref{r1}) to compare  different predictors. On the other hand, it
is shown   \cite{Ry1}  that there exists
 a predictor $R$, such that the following Cesaro average $
  t^{-1}\: \sum_{i=1}^t
\rho_{R,P} (x_1\cdots x_i) $
 goes to 0 (with probability 1) for any stationary and ergodic source
$P,$ where  $t$ goes to infinity. (This predictor will be
described in the next subsection.) That is why we will focus our
attention on such averages. From the definitions (\ref{r0}),
(\ref{r1}) and properties of the logarithm we can see that for any
probability distribution $\gamma$
$$
  t^{-1}\: \sum_{i=1}^t
\rho_{\gamma,P} (x_1\cdots x_i) =  t^{-1}\: ( \log ( P(x_1 ...
x_t) / \gamma (x_1 ... x_t) ),$$
$$ t^{-1}\: \sum_{i=1}^t
\rho^i(P\|\gamma)= t^{-1} \sum_{x_1 ... x_t \in A^t} P(x_1 ...
x_t) \log ( P(x_1 ... x_t) / \gamma (x_1 ... x_t) ).$$ Taking into
account these equations, we can see   from the definitions
(\ref{R0}) and (\ref{RR}) that the Chesaro averages of the
prediction errors (\ref{r0}) and (\ref{r1}) are equal to the
errors of estimation of limiting probabilities (\ref{R0}) and
(\ref{RR}). That is why we will use values (\ref{R0}) and
(\ref{RR}) as the main measures of the precision throughout the
chapter.

A  natural problem is to find  a 
 predictor and an estimator
 of the
limiting probabilities whose average error (\ref{RR}) is minimal
for the set of i.i.d. sources. This problem was considered and
solved by Krichevsky   \cite{Kr1,Kr}. He suggested the following
predictor:
\begin{equation}\label{kr1}
K_0(a|x_1\cdots x_t) = (\nu_{x_1\cdots x_t}(a) +1/2 )/ (t+ |A |/2
),
\end{equation}  where, as before,  $\nu_{x_1\cdots x_t}(a)$
is  the number of occurrencies of the letter $a$ in the word $x_1\ldots x_t.$ 
We can see that the Krychevsky predictor is quite close to the
Laplace's  one (\ref{L}).

\textbf{Example.} Let $ A= \{0, 1 \}, \: x_1 ... x_5 = 01010.$
Then $K_0(x_{6}=0| 01010) = (3+1/2)/ (5+1) = 7/12, K_0(x_{6}=1 |
01010) = (2+1/2)/ (5+1) = 5/12$ and $K_0(01010) = \frac{1}{2}
\frac{1}{4} \frac{1}{2} \frac{3}{8} \frac{1}{2}
 = \frac{3}{256}.$

 The Krichevsky measure  $K_0$ can be represented
 as follows:
\begin{equation}\label{Kp}
K_0(x_1 ... x_t) =\prod_{i=1}^{t} \frac{ \nu_{x_1 ... x_{i-1}}
(x_i)+ 1/2 }{i-1+|A|/2} = \frac{\prod_{a \in A} (\prod_{j=
1}^{\nu_{x_1 ... x_t}(a)} (j- 1/2))}{\prod_{i= 0}^{t-1}(i+ |A|/2)}
\,.
\end{equation}
It is known that
\begin{equation}\label{gamma}
 (r+1/2) ((r + 1) + 1/2) ... (s - 1/2) =
\frac{\Gamma(s+1/2)}{\Gamma(r+1/2)},
\end{equation}
where $\Gamma(\:)$ is the gamma function
 \cite{Kn}.
 So, (\ref{Kp}) can be presented as follows:
 \begin{equation}\label{Kp1}
K_0(x_1 ... x_t) = \: \frac{\prod_{a \in A} ( \Gamma(\nu_{x_1 ...
x_t}(a)+ 1/2 ) \,  / \Gamma(1/2) \,)   }{\Gamma(t+ |A|/2 ) \, /
\Gamma(|A|/2)  } \,.
\end{equation}
The following claim shows that the error of the Krichevsky
estimator is a half of the Laplace's one. 

\begin{claim}\label{c2}  For any i.i.d.   source $P$ generating
letters from a finite alphabet $A$ the average error (\ref{RR})
 of the estimator $K_0$  is upper bounded as
follows: \begin{equation}\label{kr2} \bar{\rho}_t (K_0, P) \equiv
\: t^{-1} \sum_{x_1 ... x_t \in A^t} P(x_1 ... x_t) \log (P(x_1
... x_t) / K_0(x_1 ... x_t))  \equiv $$ $$ t^{-1} \sum_{x_1 ...
x_t \in A^t} P(x_1 ... x_t) \log (1 / K_0(x_1 ... x_t)) - h_0(p)
\leq ((|A| - 1) \log t +C )/ (2 t) ,
\end{equation} where $C$ is a constant.
\end{claim}
Moreover, in a certain sense this average error is   minimal: it
is shown by Krichevsky \cite{Kr1} that  for any predictor $\gamma$
there exists such a source $P^*$ that
$$  \bar{\rho}_t (\gamma, P^*)  \geq  
 ((|A| - 1) \log t +C' )/ (2
t).$$ Hence, the bound $((|A| - 1) \log t +C )/ (2 t)$ cannot be 
reduced and the  Krichevsky estimator is the best (up to $O(1/t)$)
if the error is measured by the KL divergence~$\rho$.

\subsection{ Consistent Estimations and On-line Predictors  for Markov and Stationary Ergodic Processes}
Now we briefly describe consistent estimations of unknown
probabilities and efficient on-line predictors  for general
stochastic processes (or sources of information).

 First we give a formal definition of stationary ergodic processes.
  The time shift $T$ on $A^\infty$ is
defined as $T(x_1,x_2,x_3,\dots)=(x_2,x_3,\dots)$. A process $P$
is called stationary if it is $T$-invariant: $P(T^{-1}B)=P(B)$ for
every Borel set  $B\subset A^\infty$. A stationary process is
called ergodic if every $T$-invariant set has probability 0 or 1:
$P(B)=0$ or $1$ whenever $T^{-1}B=B$  \cite{Billingsley, Ga}.

We denote by $M_\infty(A)$  the set of all stationary and ergodic
sources  and let $M_0(A) \subset M_\infty(A)$ be the set of all
i.i.d. processes. We denote by  $M_m(A) \subset M_\infty(A)$ the
set of Markov sources of order (or with memory, or connectivity)
not larger than $m, \, m \geq 0.$  By definition $\mu \in M_m(A)$
if
\begin{equation}\label{ma}
\mu (x_{t+1} = a_{i_1} | x_{t} = a_{i_2}, x_{t-1} = a_{i_3},\,
...\,, x_{t-m+1} = a_{i_{m+1}},... ) \end{equation} $$ = \mu
(x_{t+1} = a_{i_1} | x_{t}  = a_{i_2}, x_{t-1} = a_{i_3},\, ...\,,
x_{t-m+1} = a_{i_{m+1}}) $$ for all $t \geq m $ and $a_{i_1},
a_{i_2}, \ldots \, \in A.$ Let $M^*(A) = \bigcup_{i=0}^\infty
M_i(A)$ be the set of all finite-order sources.

 The Laplace and
Krichevsky predictors can be extended to general Markov processes.
  The trick
is to view a Markov source $p\in M_m(A)$ as resulting from $|A|^m$
i.i.d. sources. We illustrate this idea by an example \cite{RT}.
So assume that $A=\{O,I\}$, $m=2$ and assume that the source $p\in
M_2(A)$ has generated the sequence
$$ OOIOIIOOIIIOIO. $$
We represent this sequence by the following four subsequences:
$$ **I*****I***** ,$$
$$ ***O*I***I***O ,$$
$$ ****I**O****I* ,$$
$$ ******O***IO** .$$
These four subsequences contain letters which follow $OO$,  $OI$,
$IO$ and  $II$, respectively.
 By definition,
$p\in M_m(A)$ if $p(a |x_t\cdots x_1) = p(a| x_t\cdots x_{t-m+1}
$), for all $0 < m \leq t $, all $a\in A$ and all $x_1\cdots x_t
 \in A^t $. Therefore, each of the four generated subsequences
may be considered to be generated by an i.i.d. source. Further, it
is possible to reconstruct the original sequence if we know the
four ($=|A|^m$) subsequences and the two ($=m$) first letters of
the original sequence.

Any predictor $\gamma $ for i.i.d. sources can be applied to
Markov sources. Indeed, in order to predict, it is enough to store
in the memory $|A|^m$ sequences, one corresponding to each word in
$A^m$. Thus, in the example, the letter $x_3$ which follows $OO$
is predicted based on the i.i.d. method $\gamma$ corresponding to
the $x_1 x_2$- subsequence ($=OO$),
 then $x_4$ is predicted based on the i.i.d.
method corresponding to $x_2x_3$, i.e. to the $OI$- subsequence,
and so forth. When this scheme is applied along with either  $L_0$
or  $K_0$ we denote the obtained predictors as $L_m$ and $K_m,$
correspondingly, and define the probabilities for the first $m$
letters as follows: $ L_m(x_1) = L_m(x_2) = \ldots = L_m (x_m) =
1/|A|\,, $ $K_m(x_1) = K_m(x_2) = \ldots = K_m(x_m) = 1/|A|\,. $
For example, having taken into account
 (\ref{Kp1}), we can present the Krichevsky predictors for $M_m(A)$ as follows:
\begin{equation}\label{km}
K_m(x_1 ... x_t) = \begin{cases} \frac{1}{|A|^t},& if \;\:t \leq
m\, , \cr
 & \cr
\frac {1} {|A|^{m} }  \,  \prod_{v \in A^m} \frac{\prod_{a \in
A}\:
           (( \Gamma( \nu_x(v a )+ 1/2)  \, / \, \Gamma(1/2)) }
{( \Gamma( \bar{\nu}_x(v  )+|A|/2) \,  / \,  \Gamma(|A|/2) )}, &
if \;\: t > m  \, ,\end{cases}
\end{equation}
where $\bar{\nu}_x(v  )= \sum_{a \in A} \nu_x(v a ), \:x = x_1 ...
x_t. $ It is worth noting that the representation (\ref{Kp}) can
be more convenient for carrying out calculations if $t$ is small.

\textbf{Example.}
 For the word $ OOIOIIOOIIIOIO $
considered in the previous example,    we obtain $
K_2(OOIOIIOOIIIOIO) = $ $ 2^{-2}   \,\, \,   \frac{1} {2} \frac{3}
{4}
 \,\,  \frac{1} {2} \frac{1} {4} \frac{1} {2} \frac{3} {8}  \, \, \,
 \frac{1} {2} \frac{1} {4} \frac{1} {2} \, \, \,
 \frac{1} {2} \frac{1} {4} \frac{1} {2} \, \, .
$ Here groups of multipliers   correspond to subsequences $II,$
$OIIO,$ $IOI,$ $OIO.$

In order to estimate the error of the Krichevsky predictor $K_m$
we need a general definition of the Shannon  entropy. Let $P$ be a
stationary and ergodic source generating letters from a finite
alphabet $A$. The $m-$ order (conditional) Shannon entropy and the
limiting Shannon entropy are  defined as follows:
\begin{equation}\label{moe} h_m(P) =
\sum_{v \in A^m} P(v) \sum_{a \in A} P(a/v) \log P(a/v),\qquad
 \: h_\infty(\tau) = \lim_{m \rightarrow \infty} h_m(P). \end{equation}
 (If $m=0$ we obtain the definition (\ref{h0}).) It is also known that for any $m$
\begin{equation}\label{hlim}
h_\infty(P) \leq h_m(P) \: \end{equation} \cite{Billingsley, Ga}.
\begin{claim}\label{kmc}   For any stationary and ergodic  source $P$
generating  letters from a finite alphabet $A$  the average error
 of the Krichevsky  predictor $K_m$  is upper bounded as
follows:
\begin{equation}\label{c3} -  t^{-1} \sum_{x_1 ... x_t \in A^t} P(x_1 ... x_t)
\log ( K_m (x_1 ... x_t)) - h_m(P) \leq \frac{|A|^m (|A| - 1) \log
t + C} {2 t} , \end{equation} where $C$ is a constant.
\end{claim}
 The following so-called empirical Shannon entropy, which  is an
estimation of the entropy (\ref{moe}), will play a key role in the
hypothesis testing. It will be convenient to consider its
definition  here, because this notation will be used in the proof
of the next claims.  Let $v= v_1 ... v_k$ and $x = x_1 x_2 \ldots
x_t$ be words from
  $A^*.$ Denote the rate of a word $v$ occurring in
the sequence $x = x_1 x_2 \ldots x_k$ , $x_2x_3 \ldots x_{k+1}$,
$x_3x_4 \ldots x_{k+2}$, $ \ldots $, $x_{t-k+1} \ldots x_t$ as
$\nu_x(v)$. For example, if $x = 000100$ and $v= 00, $ then $
\nu_x(00) = 3$. For any $0  \leq k < t$ the  empirical Shannon
entropy of order $k$ is defined as follows:
\begin{equation}\label{He}
h^*_{ k}( x) = -  \sum_{v \in A^k} \frac{\bar{\nu}_x(v)}{(t-k)}
\sum_{a \in A} \frac{\nu_x(va)}{ \bar{\nu}_x(v)} \log
\frac{\nu_x(va)}{ \bar{\nu}_x(v)}\, ,
\end{equation}
 where $x = x_1 \ldots x_t,$ $ \bar{\nu}_x(v  )= \sum_{a \in A} \nu_x(v a ). $
In particular, if $k=0$, we obtain $ h^*_{ 0}( x) = - t^{-1}
\sum_{a \in A} \nu_x(a)  \log (\nu_x(a) / t )\, . $

Let us define the measure $R,$  which, in fact, is a consistent
estimator of probabilities for the class of all stationary and
ergodic processes with a finite alphabet. First we define
 a probability distribution $\{\omega =
\omega_1, \omega_2, ... \}$ on integers $\{ 1, 2, ... \}$ by
\begin{equation}\label{om} \omega_1 = 1 - 1/ \log 3,\: ... \,,\:
\omega_i\,= 1/ \log (i+1) - 1/ \log (i+ 2),\: ... \; .
\end{equation}
(In what follows we will use this distribution, but results
described below are obviously true for   any distribution with
nonzero probabilities.) The measure $R$
is defined as follows:
\begin{equation}\label{R}
 R(x_1 ... x_t) = \sum_{i=0}^\infty \, \omega_{i+1} \:K_i(x_1 ...
 x_t) . \end{equation}
 It is worth
 noting that this construction can be applied to the Laplace
 measure
 (if we use $L_i$ instead of $K_i$) and any other family of measures.

\textbf{Example.}  Let us calculate $R(00), ... , R(11).$ From
(\ref{Kp}) and (\ref{km}) we obtain:$$K_0(00)= K_0(11)=
\frac{1/2}{1}\, \frac{3/2}{1+1}= 3/8,\,\: K_0(01)=
K_0(10)=\frac{1/2}{1+0} \,\frac{1/2}{1+1}= 1/8,  $$ $$ K_i(00)=
K_i(01)=K_i(10)= K_i(11)= 1/4; \:, \, i \geq 1. $$ Having taken
into account the definitions of $\omega_i$ (\ref{om}) and the
measure $R$ (\ref{R}), we can calculate $R(z_1z_2)$ as follows:
$$ R(00) = \omega_1 K_0(00) + \omega_2 K_1(00) + \ldots = (1- 1/ \log 3)\, 3/8 +(1/ \log 3- 1/ \log 4)\, 1/4 + $$
$$(1/ \log 4- 1/ \log 5)\, 1/4 + \,\ldots = (1- 1/ \log 3)\,\: 3/8 + (1/ \log 3 )\: \:1/4 \approx 0.296 .$$
Analogously, $R(01)= R(10)\approx 0.204, R(11)\approx 0.296. $

 The main properties of the measure $R$  are connected with the
Shannon entropy (\ref{moe}).

\begin{theorem}[\cite{Ry1}]\label{TR}
  For any stationary
and ergodic source $P$  the following equalities are valid:
$$  i) \, \,    \lim_{t\rightarrow\infty}
\frac {1}{t}    \log (1/ R(x_1 \cdots x_t) )     \: =  \,
h_\infty(P)
$$    with probability 1,
$$ ii) \, \, \, \,   \lim_{t\rightarrow\infty}
\frac {1}{t}  \sum_{u \in A^t} P(u)  \log ( 1 / R (u ))   \: =  \,
h_\infty(P) .
$$
\end{theorem}
So, if one uses the measure $R$ for data compression in such a way
that the codeword length of the sequence $x_1 \cdots x_t$ is
(approximately) equal to $\log (1/ R(x_1 \cdots x_t))$ bits,
he/she obtains the best achievable data compression ratio
$h_\infty(P)$ per letter. On the other hand, we know that the
redundancy of a universal code and the error of corresponding
predictor are equal. Hence, if one uses the measure $R$ for
estimation and/or prediction, the error (per letter) will go to
zero.

\subsection{ Hypothesis Testing}
Here we briefly describe the main notions of hypothesis testing
and the two particular problems considered below.
  A statistical test is formulated to
test a specific null hypothesis ($H_0$). Associated with this null
hypothesis is the alternative hypothesis ($H_1$) \cite{rng}. For
example, we will consider the two following problems:
goodness-of-fit testing (or identity testing)  and   testing of
serial independence. Both  problems are well known in mathematical
statistics and there is an extensive literature dealing with their
nonparametric testing \cite{Bab,CS,CS1,fin}.

The goodness-of-fit testing  is described as follows: a hypothesis
$H_0^{id}$ is that the source has a particular distribution $\pi$
and the alternative hypothesis $H_1^{id}$ that the sequence is
generated by a stationary and ergodic source which differs from
the source under $H_0^{id}$. One particular case, mentioned in
Introduction, is  when the source alphabet $A$ is $\{ 0, 1 \} $
and the main hypothesis $H_0^{id}$ is that a bit sequence is
generated by the Bernoulli i.i.d.
source with equal probabilities of 0's and 1's. In all cases, 
the testing should be based on a sample $x_1 \ldots x_t$ generated
by the source.

 The second problem is as follows: the null hypothesis  $H_0^{SI}$ is
that the source is Markovian of order   not larger than $m,\: (m
\geq 0),$ and the alternative hypothesis $H_1^{SI}$ is that the
sequence is generated by a stationary and ergodic source which
differs from the source under $H_0^{SI}$. In particular, if $m=0,$
this is the problem of testing for independence of time series.

For each applied test, a decision is derived that accepts or
rejects the null hypothesis. During the test, a test statistic
value is computed on the data (the sequence being tested). This
test statistic value is compared to the critical value. If the
test statistic value exceeds the critical value, the null
hypothesis is rejected. Otherwise, the null hypothesis   is
accepted. So, statistical hypothesis testing is a
conclusion-generation procedure that has two possible outcomes:
 either accept $H_0$  or accept $H_1$.

Errors of the two following types are possible: The Type I error
occurs if $H_0$ is true but the test accepts $H_1$ and, vice
versa,
 the
Type II error occurs if $H_1$ is true, but the test accepts $H_0.$
 The probability of  Type I error is
often called the level of significance of the test. This
probability can be set prior to the testing
 and is denoted  $\alpha.$
For a test, $\alpha$  is the probability that the test will say
that  $H_0$ is not true when it really is true. Common values of
$\alpha$  are about 0.01. The probabilities of  Type I and  Type
II errors  are related to each other and to the size $n$ of the
tested sequence in such a way that if two of them are specified,
the third value is automatically determined. Practitioners usually
select a sample size $n$ and a value for the probability of the
Type I error - the level of significance \cite{rng}.

\subsection{Codes}
We briefly describe the main definitions and properties (without
proofs) of lossless codes, or methods of (lossless) data
compression.
 A data compression method (or code) $\varphi$ is defined as a
set of mappings $\varphi_n $ such that $\varphi_n : A^n
\rightarrow \{ 0,1 \}^*,\, n= 1,2, \ldots\, $ and for each pair of
different words $x,y \in A^n \:$ $\varphi_n(x) \neq \varphi_n(y)
.$  It is also required that each sequence
$\varphi_n(u_1)\varphi_n(u_2) ...\varphi_n(u_r),$ $ r \geq 1,$ of
encoded words from the set $A^n, n\geq 1,$ could be uniquely
decoded into $u_1u_2 ...u_r$. Such codes are called uniquely
decodable. For example, let $A=\{a,b\}$, the code $\psi_1(a) = 0,
\psi_1(b) = 00, $  obviously, is not uniquely decodable. In what
follows we call
 uniquely decodable codes just "codes".
It is well known
 that if   $\varphi$ is a code
then  the lengths of the codewords satisfy the following
inequality (Kraft's inequality) \cite{Ga} : $ \Sigma_{u \in A^n}\:
2^{- |\varphi_n (u) |} \leq 1\:$.
 It will be
convenient to reformulate this property as follows:
\begin{claim}\label{craft}  Let $\varphi$ be a code
over an alphabet $A$. Then  for any integer $n$ there exists a
measure $\mu_\varphi$ on $A^n$ such that
\begin{equation}\label{kra}
 - \log \mu_\varphi (u) \:  \leq  \: |\varphi (u)|
\end{equation} for any $u$ from $A^n \,.$
\end{claim}
(Obviously, this claim  is true for the measure $\mu_\varphi (u)
=$ $\frac{  2^{- |\varphi (u) |} } { \Sigma_{u \in A^n}\: 2^{-
|\varphi (u)|} }).$

It was mentioned above   that, in a certain sense, the opposite
claim is true, too.  Namely, for any probability measure $\mu$
defined on $A^n, n \geq 1,$ there exists a code $\varphi_\mu$ such
that
\begin{equation}\label{opkra}
|\varphi_\mu (u) | = - \log \mu(u).
\end{equation}
(More precisely, for any $\varepsilon > 0 $  one can construct
such  a  code $\varphi_\mu ^*,$ that  $ |\varphi_\mu ^*(u) | < -
\log \mu(u) + \varepsilon $ for any $u \in A^n.$ Such a code can
be constructed by applying a so-called arithmetic coding
\cite{Ri1} .) For example, for the above described measure $R$ we
can construct a code $R_{code}$ such that
\begin{equation}\label{rcode}
|R_{code} (u) | = - \log R(u).
\end{equation}
  As we mentioned above there exist
  universal codes. For their description we recall that   sequences $x_1 \dots x_t,$
generated by a  source $P,$ can be "compressed" to the length $-
\log
 P(x_1 ... x_t)$ bits (see (\ref{opkra})) and, on the other hand, for any source $P$ there is no code $\psi$ for
 which the average codeword
 length (\,$  \Sigma
 _{u \in A^t}
 \, P(u) |\psi(u)|\,)$
 is less than  $ - \Sigma_{u \in A^t} \, P(u)\log
 P(u)$.
 Universal 
 codes can reach the lower bound
 $-
\log
 P(x_1 ... x_t)$ asymptotically for any stationary and ergodic source $P$ in average and with  probability 1.
  The formal
definition is as follows:
 a code $U$ is universal if for any
stationary and ergodic source $ P$ the following equalities are
valid:
\begin{equation}\label{U1}
\lim_{t\rightarrow\infty}  |U(x_1 \ldots x_t) |  /t = h_\infty(P)
\end{equation}  with probability 1, and
\begin{equation}\label{U2}
\lim_{t\rightarrow\infty}   E ( |U(x_1 \ldots x_t) |  ) /t =
h_\infty(P) ,
\end{equation}   where $E( f)$  is the expected  value of  $f,$ $h_\infty(P)$ is the Shannon entropy of
$P,$ see (\ref{hlim}). So, informally speaking, a universal
 code estimates the probability characteristics of a source   and
 uses  them for efficient "compression".

 In this chapter we mainly consider finite-alphabet and real-valued
 sources, but sources with countable alphabet also were considered
 by many authors \cite{BG,tcs,Ki1,RA2,RA3}. In particular, it is shown  that, for infinite alphabet, without any
condition on the source distribution it is impossible to have
universal source code and/or universal predictor, i.e. such a
predictor whose average error goes to zero, when the length of a
sequence goes to infinity. On the other hand, there are some
necessary and sufficient conditions for existence of universal
codes and predictors \cite{BG,Ki1,RA2}.

\section{Finite Alphabet Processes}\label{s:21}
\subsection{ The Estimation  of (Limiting) Probabilities }

The following theorem shows how universal codes can be applied for
probability estimation. 
\begin{theorem}\label{KorotkoObozvatEtuTeoremu}   Let $U$ be a universal code and
\begin{equation}\label{UC}    \mu_U (u) =
2^{- |U (u) |} / \Sigma_{ v \in A^{|u|}}\: 2^{- |U(v)|} .
\end{equation}
Then, for any stationary and ergodic source $P$  the following
equalities are valid:
$$  i) \, \,    \lim_{t\rightarrow\infty}
\frac {1}{t} ( -   \log  P(x_1 \cdots x_t) -  (-  \log  \mu_U(x_1
\cdots x_t)  ) ) \: =  \, 0
$$   with probability 1,
$$ ii) \, \, \, \,   \lim_{t\rightarrow\infty}
\frac {1}{t}  \sum_{u \in A^t} P(u)  \log ( P(u) / \mu_U (u )) \:
= \, 0 .
$$
\end{theorem}
The informal outline of the proof ia as follows: $\frac {1}{t} ( -
\log P(x_1 \cdots x_t)$ and $\frac {1}{t} (- \log  \mu_U(x_1
\cdots x_t)  )$ goes to the Shannon entropy $h_\infty(P),$ that is
why the difference is 0.

 So, we can see that, in a
certain sense, the measure $\mu_U$ is a consistent nonparametric
estimation of the (unknown)  measure $P.$

Nowadays there are many efficient universal codes (and universal
 predictors connected with them),  which can be applied to estimation.
For example, the above described measure $R$ is based on a
universal code  \cite{Ry0,Ry1} and can be applied for probability
estimation. More precisely, Theorem~\ref{KorotkoObozvatEtuTeoremu}
 (and the following theorems)
 are true
for $R$, if we replace $\mu_U$ by $R.$

 It is important to note that the  measure $R$ has some additional
properties, which can be useful for  applications. The following
theorem describes these properties (whereas all other theorems are
valid for all universal codes and corresponding  measures,
including the measure $R ).$
\begin{theorem}\label{t3}(\cite{Ry0,Ry1} )
For any Markov process $P$ with memory $k$
\begin{itemize}
\item[i)] the error of the probability estimator, which is based
on the measure $R,$ is upper-bounded as follows:
$$   \frac {1}{t}  \sum_{u \in A^t} P(u)  \log ( P(u) / R(u ))   \: \leq   \frac {(|A|-1) |A|^{k}  \log t} {2\,t}
+ O\left(\frac {1}{t}  \right),
$$
\item[ii)]  the error of $R$ is asymptotically minimal in the following sense: for any measure $\mu$ 
there exists a $k-$memory Markov process $p_\mu$ such that
$$   \frac {1}{t}  \sum_{u \in A^t} p_\mu(u)  \log ( p_\mu(u) / \mu(u ))
 \geq  \frac {\left(|A|-1\right) |A|^{k}  \log t} {2\,t} + O\left(\frac {1}{t}\right),
$$
\item[iii)]Let $\Theta$ be  a set of stationary and ergodic
processes such that there  exists a measure $\mu_\Theta$ for which
the estimation error of the probability goes to 0 uniformly:
$$  \lim_{t\rightarrow\infty} \; \sup_{P \in \Theta} \left( \frac {1}{t}
 \sum_{u \in A^t} P(u)  \log \left( P(u) / \mu_\Theta(u )\right) \right)= 0.$$
Then the error of the estimator which is based on the measure $R,$
goes to 0 uniformly too:
$$  \lim_{t\rightarrow\infty} \; \sup_{P \in \Theta} \left(\frac {1}{t}
 \sum_{u \in A^t} P(u)  \log ( P(u) / R(u ))\right)= 0.$$
\end{itemize}
\end{theorem}

\subsection{ Prediction}

As we mentioned above, any universal code $U$ can be applied for
prediction. Namely, the  measure $\mu_U$   (\ref{UC}) can be used
for prediction as the following conditional probability:
\begin{equation}\label{preu}
\mu_U(x_{t+1}| x_1 ... x_t) =  \mu_U(x_1 ... x_t x_{t+1})/ \mu_U(
x_1 ... x_t).
\end{equation}
 The following theorem shows that such a
predictor is quite reasonable. Moreover, it gives a possibility to
apply practically used data compressors for prediction of real
data (like EUR/USD rate)  and obtain quite precise estimation 
\cite{RM1} .
\begin{theorem}\label{t4}  Let $U$ be a universal code and $P$
be any stationary and ergodic process. Then
$$
i)  \,   \lim_{t\rightarrow\infty}   \frac {1}{t}   \, \, E \{
\log \frac{ P(x_1) } { \mu_U(x_1) } + \log  \frac{ P(x_2|x_1) } {
\mu_U(x_2|x_1) }+  \ldots + \log  \frac{ P(x_{t}|x_1 ...
 x_{t-1}) } { \mu_U(x_{t}|x_1 ... x_{t-1}) }
 \} \: = \,0, $$
$$ ii)   \,   \lim_{t\rightarrow\infty} E(  \frac {1}{t}   \, \,
\sum_{i=0}^{t-1} ( P(x_{i+1}|x_1 ... x_i)  - \mu_U(x_{i+1}|x_1 ...
x_i) )^2 )\:=\,0\,,$$ and
$$ iii)   \,   \lim_{t\rightarrow\infty} E(  \frac {1}{t}   \, \,
\sum_{i=0}^{t-1} | P(x_{i+1}|x_1 ... x_i)  - \mu_U(x_{i+1}|x_1 ...
x_i) | )\:=\,0\,.$$  \end{theorem} An informal outline of the
proof is as follows:
$$\frac {1}{t}   \, \, \{E(
\log  \frac{ P(x_1) } { \mu_U(x_1) }) + E(\log  \frac{ P(x_2|x_1)
} { \mu_U(x_2|x_1) } )+  \ldots + E(\log  \frac{ P(x_{t}|x_1 ...
 x_{t-1}) } { \mu_U(x_{t}|x_1 ... x_{t-1}) })
\}$$ is equal to $\frac {1}{t} E(\log  \frac{ P(x_1 ...
 x_{t}) } { \mu_U( x_1 ... x_{t}) } ).$ Taking into account
 Theorem~\ref{KorotkoObozvatEtuTeoremu}, we obtain the first
 statement of the theorem.

\textbf{Comment 1.} The  measure $R$ described above has one
additional property if it is used for prediction. Namely, for any
Markov process $P$ ($P \in M^*(A))$  the following is true:
$$  \,   \lim_{t\rightarrow\infty}     \, \,\log  \frac{ P(x_{t+1}|x_1 ...
 x_t) } { R(x_{t+1}|x_1 ... x_t)}
=0 $$ with probability 1, where $R(x_{t+1}|x_1 ... x_t)=R(x_1 ...
x_tx_{t+1})/R(x_1 ... x_t)$   \cite{Ry2}.

\textbf{Comment 2.}   It is known \cite{DR}   that, in fact, the
statements ii) and iii) are equivalent.

\subsection{ Problems with Side Information }
Now we consider the  so-called problems with side information,
which are described as follows: there is a stationary and ergodic
source whose
 alphabet $A$ is presented as a product $ A = X \times Y.$ We are given a sequence
 $  (x_1,y_1), \ldots, $ $ (x_{t-1},y_{t-1})$ and
side information $y_t.$ The goal is to predict, or estimate,
$x_t.$
 This problem arises in statistical decision theory, pattern recognition, and machine learning.
   Obviously, if someone knows the conditional probabilities $P(x_t| $ $(x_1,y_1), \ldots, $ $ (x_{t-1},y_{t-1}), y_t)$
 for all $x_t \in X, $ he has all  information  about $x_t,$ available before $x_t$ is known.
 That is why  we will look for the best (or, at least, good) estimations for this conditional probabilities.
 Our solution will be based on results obtained in the previous subsection.
 More precisely, for any
 universal code $U$ and the corresponding measure $\mu_U$ (\ref{UC}) we define the following estimate
 for the problem with side information:$$
\mu_U(x_t| (x_1,y_1), \ldots, (x_{t-1},y_{t-1}), y_t) =
$$ $$
\frac{\mu_U( (x_1,y_1), \ldots, (x_{t-1},y_{t-1}), (x_t,y_t)) }
{\sum_{x_t \in X}\mu_U( (x_1,y_1), \ldots, (x_{t-1},y_{t-1}),
(x_t,y_t))}.$$ The following theorem shows that this estimate is
quite reasonable.

\begin{theorem}\label{t5}  Let $U$ be a universal code and let 
$P$ be any stationary and ergodic process. Then
$$
i)  \,   \lim_{t\rightarrow\infty}   \frac {1}{t}   \, \, \{E(
\log  \frac{ P(x_1|y_1) } { \mu_U(x_1|y_1) }) + E(\log  \frac{
P(x_2|(x_1,y_1),y_2) } { \mu_U(x_2|(x_1,y_1),y_2) }) +  \ldots $$
$$+ E(\log  \frac{ P(x_{t}|(x_1,y_1), ..., (x_{t-1},y_{t-1}), y_t)
} { \mu_U(x_{t}|(x_1,y_1), ..., (x_{t-1},y_{t-1}), y_t) }) \} \: =
\,0 ,$$
$$ ii)   \,   \lim_{t\rightarrow\infty} E(  \frac {1}{t}   \, \,
\sum_{i=0}^{t-1} ( P(x_{i+1}|(x_1,y_1), ..., (x_{i},y_{i}),
y_{i+1}))  - $$ $$ \mu_U(x_{i+1}|(x_1,y_1), ..., (x_{i},y_{i}),
y_{i+1}))^2 )\: \quad=\,0\,,$$  and
$$ iii)   \,   \lim_{t\rightarrow\infty} E(  \frac {1}{t}   \, \,
\sum_{i=0}^{t-1} | P(x_{i+1}|(x_1,y_1), ..., (x_{i},y_{i}),
y_{i+1}))  - $$ $$ \mu_U(x_{i+1}|(x_1,y_1), ..., (x_{i},y_{i}),
y_{i+1})| )\: \quad=\,0\,.$$  \end{theorem} The proof is very
close to the proof of the previous theorem.

\subsection{ The Case of Several Independent Samples   }
In this part we consider a situation which is important for
practical applications, but needs cumbersome notations. Namely,
 we extend our consideration to the case where the sample is
presented as several independent samples
 $x^1= x_1^1 \ldots
x_{t_1}^1,$ $x^2 = x_1^2 \ldots x_{t_2}^2, ... ,$ $x^r = x_1^r
\ldots x_{t_r}^r $ generated by a source. More precisely, we will
suppose that all sequences were independently  created by one
stationary and
ergodic source. (The point is that  it is impossible 
 just to combine all samples into one, if the source  is not i.i.d.) We
denote them  by $x^1 \diamond x^2 \diamond \ldots \diamond x^r$
and define
 $\nu_{x^1\diamond
x^2\diamond ... \diamond x^r } (v) = \sum_{i=1}^r \nu_{x^i}(v) .$
For example, if $x^1 = 0010, x^2 = 011,$ then $\nu_{x^1\diamond
x^2}(00)= 1.$ The definition of  $K_m$ and $R$ can be extended to
this case:
\begin{equation}\label{km1}
K_m(x^1\diamond x^2\diamond ... \diamond x^r  ) =  \end{equation}
 $$ (
\prod_{i=1}^r |A|^{\, -  \min{ \{m, t_i\}    } }  \,  ) \, \, \,
\, \prod_{v \in A^m} \frac{\prod_{a \in A}\:
           (( \Gamma( \nu_{x^1\diamond x^2\diamond ... \diamond x^r }(v a )+ 1/2)  \, / \, \Gamma(1/2)) }
{( \Gamma( \bar{\nu}_{x^1\diamond x^2\diamond ... \diamond x^r }(v
)+|A|/2) \,  / \,  \Gamma(|A|/2) )},  \,$$ whereas the definition
of $R$  is the same (see (\ref{R}) ). (Here, as before,
$\bar{\nu}_{x^1\diamond x^2\diamond ... \diamond x^r }(v  )  =$ $
\sum_{a \in A} \nu_{x^1\diamond x^2\diamond ... \diamond x^r }(v a
). $ Note, that $\bar{\nu}_{x^1\diamond x^2\diamond ... \diamond
x^r }(\,)=\sum_{i=1}^r t_i$ if $m=0$.)

The following example is intended to show the difference between
the case  of many samples and one.

\textbf{Example.} Let there be two independent samples $y= y_1
\ldots y_4 = 0101$ and $x = x_1 \ldots x_3 = 101,$ generated by a
stationary and ergodic source with the alphabet $\{0,1\}.$ One
wants to estimate the (limiting) probabilities $P(z_1z_2), z_1,z_2
\in \{0,1\} $ (here $z_1z_2 \ldots $ can be considered as an
independent sequence, generated by the source) and predict
$x_4x_5$  (i.e. estimate conditional probability $P(x_4x_5|x_1
\ldots x_3 = 101, y_1 \ldots y_4 = 0101).$ For solving both
problems we will use the measure $R$ (see (\ref{R})). First we
consider the case where $P(z_1z_2)$ is to be estimated without
knowledge of  sequences $x$ and $y.$ Those probabilities were
calculated previously  and we obtained: $ R(00) \approx 0.296,$ $
R(01)= R(10)\approx 0.204,$ $ R(11)\approx 0.296.$ Let us now
estimate the probability $P(z_1z_2)$ taking into account that
there are two independent samples $y= y_1 \ldots y_4 = 0101$ and
$x = x_1 \ldots x_3 = 101.$ First of all we note that such
estimates are based on the formula for conditional probabilities:
$$ R(z|x\diamond y) = R(x\diamond y \diamond z)/R(x\diamond y) .$$
Then we  estimate the frequencies: $ \nu_{\,0101\diamond 101
}(0)=3,$ $  \nu_{\,0101\diamond 101  }(1)=4,$  $\:$ $
\nu_{\,0101\diamond 101 }(00) $ $= \nu_{\,0101\diamond 101 }(11)=0
,$ $ \nu_{\,0101\diamond 101 }(01)= 3,$ $ \nu_{\,0101\diamond 101
}(10)=2,$ $ \nu_{\,0101\diamond 101 }(010)=1,$ $
\nu_{\,0101\diamond 101 }(101)=2,$ $ \nu_{\,0101\diamond 101
}(0101)=1, $ whereas frequencies of all other three-letters and
four-letters words are 0. Then we calculate :
$$ K_0(\,0101\diamond 101 ) = \frac{1}{2} \frac{3}{4} \frac{5}{6}\frac{7}{8}\frac{1}{10}\frac{3}{12}\frac{5}{14}
\approx 0.00244,\,$$ $$ K_1(\,0101\diamond 101 ) = (2^{-1})^2
\:\frac{1}{2} \frac{3}{4} \frac{5}{6}\, \:1 \:\frac{1}{2}
\frac{3}{4}\, \:1   $$
$$\approx0.0293,\quad K_2(\,0101\diamond 101 ) \approx 0.01172,\quad K_i(\,0101\diamond 101 ) = 2^{-7},\: i\geq 3 ,
$$ $$
R(\,0101\diamond 101 ) = \omega_1 K_0(\,0101\diamond 101 ) +
\omega_2 K_1(\,0101\diamond 101 ) + \:\ldots \approx
$$ $$
0.369 \:0.00244 \,+ \, 0.131 \: 0.0293 \,+ 0.06932 \: 0.01172 \:+
2^{-7} \; /\, \log 5  \approx 0.0089. $$ In order to avoid
repetitions, we estimate only one probability $P(z_1z_2= 01).$
Carrying out similar  calculations, we obtain $ R(0101\diamond 101
\diamond 01) \approx 0.00292,$ $ R(z_1z_2= 01|y_1 \ldots y_4 =
0101,x_1 \ldots x_3 = 101) = $ $  R(0101\diamond 101 \diamond 01)/
R(\,0101\diamond 101) \approx 0.32812. $ If we compare this value
and the estimation $R(01) \approx 0.204$, which is not based on
the knowledge of samples $x$ and $y$, we can see that  the measure
$R$ uses   additional information quite naturally (indeed,
 $01$ is quite frequent in $ y = y_1 \ldots y_4 = 0101$ and $
x= x_1 \ldots x_3 = 101$).

Such generalization can be applied to
 many universal codes, but, generally speaking,
there exist codes $U$ for which $U(x^1\diamond x^2)$ is not
defined and, hence, the measure $\mu_U(x_1\diamond x_2)$ is not
defined. That is why we will  describe properties of the universal
code $R,$ but not of universal codes in general.
 For the measure $R$
all asymptotic properties are the same for the cases of one sample
and several samples.
 More precisely, the following statement is true:

\begin{claim}\label{cl2} Let $x^1,x^2, ... ,x^r $
 be  independent sequences  generated by a stationary and ergodic source
and let 
 $t$ be a total length of these 
 sequences $(t = \sum_{i=1}^r
|x^i| ).$ Then, if $t \rightarrow \infty,$ (and $r$ is fixed) the
statements of the Theorems~\ref{KorotkoObozvatEtuTeoremu} -
~\ref{t5} are valid, when applied to $x^1\diamond x^2\diamond ...
\diamond x^r $ instead of
 $x_1\ldots x_t.$ (In theorems~\ref{KorotkoObozvatEtuTeoremu} - ~\ref{t5} $\:\mu_U$ should be  
changed to $\:R$.)
\end{claim}

The proofs are completely analogous to the proofs of the
Theorems~\ref{KorotkoObozvatEtuTeoremu}---\ref{t5}. 

Now we can extend the definition of the empirical Shannon  entropy
(\ref{He}) to the 
case of several words $x^1= x_1^1 \ldots x_{t_1}^1,$ $x^2 = x_1^2
\ldots x_{t_2}^2, ... ,$ $x^r = x_1^r \ldots x_{t_r}^r. $
 We define $\nu_{x^1\diamond x^2\diamond ... \diamond x^r
} (v) = \sum_{i=1}^r \nu_{x^i}(v) .$ For example, if $x^1 = 0010,
x^2 = 011,$ then $\nu_{x^1\diamond x^2}(00)= 1.$ Analogously to
(\ref{He}),

\begin{equation}\label{He1}
h^*_{ k}( x^1\diamond x^2\diamond ... \diamond x^r ) = -  \sum_{v
\in A^k} \frac{\bar{\nu}_{x^1\diamond
 ... \diamond x^r }(v)}{(t-k r)} \sum_{a \in A}
\frac{\nu_{x^1\diamond  ... \diamond x^r }(va)}{
\bar{\nu}_{x^1\diamond  ... \diamond x^r }(v)} \log
\frac{\nu_{x^1\diamond  ... \diamond x^r }(va)}{
\bar{\nu}_{x^1\diamond  ... \diamond x^r }(v)}\, ,\end{equation}
where  $ \bar{\nu}_{x^1\diamond  ... \diamond x^r }(v  )= \sum_{a
\in A} \nu_{x^1\diamond  ... \diamond x^r }(v a ). $

For any sequence of words  $x^1= x_1^1 \ldots x_{t_1}^1,$ $x^2 =
x_1^2 \ldots x_{t_2}^2, ... ,$ $x^r = x_1^r \ldots x_{t_r}^r $
from $A^*$ and any measure $\theta$ we define $\theta (x^1
\diamond x^2 \diamond \ldots \diamond x^r) = \prod_{i=1}^r
\theta(x^i) .$ The following lemma gives an upper bound for
unknown probabilities.
\begin{lemma}\label{Le} Let $\theta$ be a measure from $M_m(A), m
\geq 0,$ and $x^1, \ldots, x^r$ be words from  $A^*,$ whose
lengths are not less than $m.$  Then
\begin{equation}\label{L}
\theta (x^1 \diamond \ldots \diamond x^r) \leq
 \, 2^{- (t- r m) \, h^*_m(x^1 \diamond
...\diamond x^t)},\end{equation} where $\theta (x^1 \diamond
\ldots \diamond x^r) = \prod_{i=1}^r \theta (x^i).$
\end{lemma}

\section{Hypothesis Testing}\label{s:22}

\subsection{Goodness-of-Fit or Identity Testing}

 Now we consider the problem of testing $H_0^{id}$ against $H_1^{id}.$
Let us recall that
 the hypothesis
$H_0^{id}$ is that the source has a particular distribution $\pi$
and the alternative hypothesis $H_1^{id}$ that the sequence is
generated by a stationary and ergodic source which differs from
the source under $H_0^{id}$.
  Let the required  level of significance (or the  Type I error)
   be $\alpha ,\, \alpha \in (0,1).$  We
describe a statistical test which can be constructed based on any
code $\varphi$.

The  main idea of the suggested test is quite natural: compress a
sample sequence $x_1... x_t$ by a code $\varphi$. If the length of
the  codeword ($|\varphi(x_1... x_t)|$) is significantly less than
the value $- \log \pi(x_1... x_t),$ then $H_0^{id}$ should be
rejected. The key observation is that the probability of all
rejected sequences is quite small for any $\varphi$, that is why
the Type I error can be made small. The precise description of the
test is as follows: { \it
 The hypothesis
$H_0^{id}$  is accepted if
\begin{equation}\label{t1} - \log
\pi(x_1... x_t) - |\varphi (x_1... x_t) | \leq - \log \alpha
.\end{equation}
  Otherwise, $H_0^{id}$ is rejected.}
We denote this test by $T_{\varphi}^{\,id}(A, \alpha).$

 \begin{theorem}\label{test1} i) For each distribution  $\pi, \alpha
\in (0,1)$ and  a code $\varphi$, the Type I error of the
described test $T_{\varphi}^{\,id}(A, \alpha)$ is not larger than
$\alpha $ and ii) if, in addition, $\pi$ is a finite-order
stationary and ergodic process over $A^\infty$ (i.e. $\pi \in
M^*(A)$) and $\varphi$ is a universal code, then the Type II error
of the test $T_{\varphi}^{\,id}(A, \alpha)$ goes to 0, when $t$
tends to infinity.
\end{theorem}

\subsection{Testing for Serial Independence} 

Let us recall that the null hypothesis  $H_0^{SI}$ is that the
source is Markovian of order   not larger than $m,\: (m \geq 0),$
and the alternative hypothesis $H_1^{SI}$ is that the sequence is
generated by a stationary and ergodic source which differs from
the source under $H_0^{SI}$. In particular, if $m=0,$ this is the
problem of testing for independence of time series.

Let there be given a sample $x_1 ... x_t $ generated by an
(unknown) source $\pi.$ The main hypothesis $H_0^{SI}$ is that the
source $\pi$ is Markovian whose
 order is not  greater than $m,\: (m \geq 0),$
and the alternative hypothesis $H_1^{SI}$ is that the sequence is
generated by a stationary and ergodic source which 
 differs from
the source under $H_0^{SI}$. The described  test is as follows.

\emph{Let $\varphi$ be any code.
 By definition, the hypothesis $H_0^{SI}$ is accepted if
\begin{equation}\label{cr}
 (t-m)\: h^*_{m}(x_1 ... x_t) -  |\varphi(x_1 ... x_t)|  \leq
\log (1 / \alpha) \,,
\end{equation} where $\alpha \in (0,1) .$  Otherwise, $H_0^{SI}$ is rejected.}
 We denote this
test by $T_{\, \varphi}^{\,SI}(A,\alpha).$

\begin{theorem}\label{test2} i) For any code $\varphi$ the Type I
error of the test $T_{\, \varphi}^{\,SI}(A,\alpha)$ is less than
or equal to $\alpha, \alpha \in (0,1)$ and, ii) if, in addition,
$\varphi$ is a universal code, then the Type II error of the test
$T_{\, \varphi}^{\,SI}(A,\alpha)$ goes to 0, when $t$ tends to
infinity. \end{theorem}

\section{ Real-Valued Time Series }\label{s:3}
\subsection{ Density Estimation and Its Application}

Here we address the  problem of nonparametric estimation of the
density for   time series. Let $X_t$ be a
 time series  and the probability distribution of $X_t$ is
unknown, but it is known that the time series  is stationary and
ergodic.
 We have seen
that Shannon-MacMillan-Breiman theorem played a key role in the 
 case of finite-alphabet processes. In this part we will use its
generalization to the processes  with densities, which was
established  by Barron \cite{Ba}.
 First we describe considered
processes with some properties needed for
the generalized Shannon-MacMillan-Breiman theorem to hold. 
In what follows, we restrict our attention to processes that take
bounded real valued.  However,  the main results  may be extended
to processes taking values in a compact subset of a  separable
metric space.

Let  $B$ denote the Borel subsets of $\textsf{R}$, and $B^k$
denote the Borel subsets of $\textsf{R}^k,$ where $\textsf{R}$ is
the set of real numbers. Let $\textsf{R}^\infty$ be the set of all
infinite sequences $x = x_1, x_2\ldots $ with $x_i \in
\textsf{R}$, and let $B^\infty$ denote the usual product sigma
field on $\textsf{R}^\infty$, generated by the finite dimensional
cylinder sets $ \{ A_1, \ldots A_k,  \textsf{R}, \textsf{R},
\ldots \}$, where $ A_i \in B, i = 1, \ldots , k.$
 Each stochastic process   $X_1, X_2, \ldots , X_i \in \textsf{R},$ is defined by a probability distribution
 on $(\textsf{R}^\infty, B^\infty)$. Suppose that the joint distribution $P_n$ for $(X_1,
X_2, \ldots, X_n)$ has a probability density function $p(x_1 x_2
\ldots x_n)$ with respect to a sigma-finite measure $M_n$. Assume
that the sequence of dominating measures $M_n$ is Markov of order
$m \geq 0$ with a stationary transition measure. A  familiar 
case for $M_n$ is Lebesgue measure. Let $p(x_{n+1}|x_1 \ldots
x_n)$ denote the conditional density
 given by the ratio $p(x_1 \ldots x_{n+1}) $ $/ p(x_1 \ldots x_n)$ for
 $n > 1.$ It is known that for stationary and ergodic processes there exists a
 so- called relative entropy rate $\tilde{h }$ defined  by
\begin{equation}\label{ent}\tilde{h} =   \lim_{n \rightarrow \infty} - E(
\log p(x_{n+1}|x_1 \ldots x_n) ), \end{equation} where $E$ denotes
expectation with respect to $P $. We will use the following
generalization of the Shannon-MacMillan-Breiman
 theorem:

\ \begin{claim}[\cite{Ba}]\label{ba1} If $\{ X_n \}$ is a
$P-$stationary ergodic process with density $p(x_1 \ldots x_n) =
dP_n/dM_n$ and $\tilde{h}_n < \infty $ for some $n\geq m,$  the
sequence of relative entropy densities $ - (1/n) \log p(x_1 \ldots
x_n)$ convergence  almost surely to the relative entropy rate,
i.e.,
\begin{equation}\label{ba}  \lim_{n\rightarrow\infty} ( - 1/n) \log p(x_1 \ldots x_n) = \tilde{h}
\end{equation}  with probability 1 (according to $P$).
\end{claim}

Now we return to the estimation problems.
Let $\{ \Pi_n \}, n \geq 1,$ be an increasing sequence of finite 
partitions of $\textsf{R}$ that asymptotically generates the Borel
sigma-field $B$  and let $x^{[k]}$ denote the element of $\Pi_k$
that contains the point $x.$ (Informally,  $x^{[k]}$ is obtained
by quantizing $x$ to $k$ bits of precision.) For integers $s$ and
$n$ we define the following approximation of the density
\begin{equation}\label{ds}
p^s (x_1 \ldots x_n) =  P(x_1^{[s]} \ldots  x_n^{[s]}) /
M_n(x_1^{[s]} \ldots x_n^{[s]}).
 \end{equation}
  We also consider
\begin{equation}\label{ents} \tilde{h}_s =   \lim_{n \rightarrow \infty} - E(
\log p^s(x_{n+1}|x_1 \ldots x_n)). \end{equation} Applying the
claim 2  to the density $p^s (x_1 \ldots x_t),$ we obtain that
a.s. 
\begin{equation}\label{Ba2}
\lim_{t\rightarrow\infty} -  \frac{1}{t} \log p^s (x_1 \ldots x_t)
= \tilde{h}_s.
\end{equation}
Let $U$ be a universal code, which is defined for any finite
alphabet. In order to describe a density estimate we will use the
probability  distribution  $\omega_i, i=1,2, \ldots ,$ see
(\ref{om}) (In what follows we will use this distribution, but
results described below are obviously true for any distribution
with nonzero probabilities.) Now we can define the density
estimate $r_U$ as follows:
\begin{equation}\label{den}    r_U(x_1 \ldots x_t) =
\sum_{i=0}^\infty \omega_i \:\mu_U(x_1^{[i]} \ldots x_t^{[i]}) /
M_t(x_1^{[i]} \ldots x_t^{[i]})\:,
 \end{equation} where  the measure $\mu_U$ is defined by
 (\ref{UC}).
 (It is assumed 
here that the code $U(x_1^{[i]} \ldots x_t^{[i]})$
 is defined for the alphabet, which contains $|\Pi_i|$ letters.)

It turns out that, in a certain sense,
 the density $r_U(x_1 \ldots x_t)$ estimates the 
 unknown density $p(x_1 \ldots  x_t).$

\begin{theorem}\label{t6}   \emph{Let $X_t$ be a stationary ergodic
process with  densities $p(x_1 \ldots  x_t)$ $ = dP_t / d M_t$
such that
\begin{equation}\label{hh} \lim_{s\rightarrow\infty} \tilde{h}_s = \tilde{h}  < \infty,\end{equation} where  $
\tilde{h}$ and $\tilde{h}_s$ are
 relative entropy rates, see  (\ref{ent}), (\ref{ents}). Then }
\begin{equation}\label{thm}
\lim_{t\rightarrow\infty}  \frac{1}{t}  \log  \frac{p(x_1 ...
x_t)}{r_U(x_1 ... x_t)}     = 0 \end{equation} \emph{with
probability 1 and
\begin{equation}\label{th}
\lim_{t\rightarrow\infty}  \frac{1}{t} \: E (\log \frac{p(x_1
\ldots x_t)}{r_U(x_1 \ldots x_t)} ) \:= \: 0 \: .
 \end{equation}}
 \end{theorem}

We have seen that the requirement (\ref{hh}) plays an important
role in the proof. The natural question is whether there exist
processes for which (\ref{hh}) is valid. The answer is positive.
For example, let a process possess   values in the  interval $
[-1,1],$ $M_n$ be Lebesgue measure and the considered process is
Markovian with conditional density

$$
p(x|y)=\begin{cases}1/2 + \alpha \: \; sign(y)    ,&if \: \:x < 0
\, \cr
                      1/2 - \alpha \:\; sign(y)  , &if \:\:x \geq 0\,
                      ,\end{cases}
$$
where $\alpha \in (0,1/2)$ is a parameter and $$
sign(y)=\begin{cases} - 1,&if\: \: y  < 0 ,\cr
                      1, &if \:\: y\geq 0\, .\end{cases}
$$
In words, the density depends on a sign of the previous value. If
the value is positive, then the density is more than 1/2,
otherwise it is less than 1/2. It is easy to see that (\ref{hh})
is true for any $\alpha \in (0,1).$

The following two theorems are devoted to the conditional
probability $r_U(x | x_1 ... x_m)$ $= r_U(x_1 ... x_m x)/ r_U( x_1
... x_m)$ which, in turn, is connected with
 the prediction problem. We will see that the conditional density $r_U(x | x_1 ...
 x_m)$ is a reasonable estimation of the unknown density $p(x | x_1 ... x_m).$

\begin{theorem}\label{t7}    Let $B_1, B_2, ... $ be a  sequence
of  measurable    sets. Then the following equalities  are true:
\begin{equation}\label{thc} i)\:
\lim_{t\rightarrow\infty}   \: E( \frac{1}{t} \sum_{m=0}^{t-1} (
P(x_{m+1} \in B_{m+1}  | x_1 ... x_m) -   R_U(x_{m+1} \in B_{m+1}
| x_1 ... x_m) )^2 ) = 0\,,
\end{equation}
$$ ii) \:
E( \frac{1}{t} \sum_{m=0}^{t-1} | P(x_{m+1} \in B_{m+1}  | x_1 ...
x_m) -   R_U(x_{m+1} \in B_{m+1}  | x_1 ... x_m) )| = 0\,,
$$
where $R_U(x_{m+1} \in B_{m+1}  | x_1 ... x_m)  = \int_{B_{m+1} }
 r_U(x | x_1 ...
 x_m) dM_{1/m}  $\end{theorem}

We have seen that in a certain sense the estimation  $r_U$
approximates the unknown density $p.$  The following theorem shows
that $r_U$ can be used instead of $p$ for estimation of average
values of certain functions.

\begin{theorem}\label{t8}  \emph{Let $f$ be   an integrable
function, whose absolute value is bounded by a certain constant
$\bar{M}$ and all conditions of the theorem 2 are true. Then the
following equality is valid:}
\begin{equation}\label{int} i) \,
\lim_{t\rightarrow\infty}   \frac{1}{t} E( \sum_{m=0}^{t-1}  (
\int  f(x) \, p(x  | x_1 ... x_m) d M_m  -  \int  f(x)\, r_U(x |
x_1 ... x_m) d M_m  )^2 )  = 0,
\end{equation} $$
ii) \, \lim_{t\rightarrow\infty}   \frac{1}{t} E( \sum_{m=0}^{t-1}
| \int  f(x)\, p(x  | x_1 ... x_m)\, d M_m  -  \int  f(x) \,r_U(x
| x_1 ... x_m) \,d M_m  | )  = 0. $$
\end{theorem}

It is worth noting that this approach was used for prediction of
real processes \cite{RM1}.

\subsection{Hypothesis Testing}

In this subsection  we consider a case where the source alphabet
$A$ is infinite, say, a part of $\textsf{R}^n$. Our strategy is to
use finite partitions of $A$ and to consider hypotheses
corresponding to the partitions. This approach can be directly
applied
 to the
goodness-of-fit testing, but it cannot be applied  to the serial
independence testing.  The point is that  if someone combines
letters (or states) of a Markov chain, the chain order (or memory)
can increase. For example, if the alphabet contains three letters,
there exists  a Markov chain of order one, such that  combining
 two letters into one  transforms the chain into 
 a process
with infinite memory.  That is why in this part we will consider
the  independence testing for i.i.d. processes only (i.e.
processes from $M_0(A)$).

In order to avoid repetitions, we will consider a general scheme,
which can be applied to both
  tests using  notations $H_0^\aleph, H_1^\aleph$
 and $T_{\varphi}^{\,\aleph}(A, \alpha),$
 where $\aleph$
is an abbreviation of one of the described tests (i.e. \emph{id }
and \emph{SI}.)

 Let us  give some definitions. Let
$\Lambda = \lambda_1, ..., \lambda_s$ be a finite (measurable)
partition of $A$ and let $\Lambda(x)$ be an element of the
partition $\Lambda$ which contains $x \in A.$ For any process
$\pi$ we define a process $\pi_\Lambda$ over a new alphabet
$\Lambda$ by the equation
$$ \pi_\Lambda( \lambda_{i_1}... \lambda_{i_k}) = \pi( x_1 \in \lambda_{i_1}, ...,
x_k \in \lambda_{i_k} ),$$ where $x_1 ... x_k \in A^k .$

 We will consider an
infinite sequence  of partitions $\hat{\Lambda}= \Lambda_1,
\Lambda_2, ....$
 and say that such a  sequence discriminates between a
 pair of hypotheses $H_0^\aleph(A), H_1^\aleph(A)$ about processes, 
  if for each process $\varrho, $ for which $H_1^\aleph(A)$ is true,
 there exists a partition $\Lambda_j$ for which $H_1^\aleph(\Lambda_j)$
 is true for the process $\varrho_{\Lambda_j}.$

Let $H_0^\aleph(A), H_1(A)^\aleph$ be a pair of hypotheses,
$\hat{\Lambda}= \Lambda_1, \Lambda_2, ...$ be a sequence of
partitions, $\alpha $ be from $ (0,1)$ and $\varphi$ be a code. 
The scheme for both tests is as follows:

 { \it
 The hypothesis
$H_0^{\aleph}(A)$  is accepted if for all $i = 1, 2, 3, ... $  the
test $T_\varphi^\aleph (\Lambda_i, (\alpha \omega_i) )$ accepts
 the hypothesis $H_0^\aleph(\Lambda_i).$
  Otherwise, $H_0^\aleph$ is rejected.} We denote this test  $\textbf{T}_{
\alpha,\varphi}^{\aleph}(\hat{\Lambda}).$

\textbf{Comment 3.}  It is important to note that one does not
need to
check an infinite number of inequalities when  
 applying this
test. The point is that the hypothesis $H_0^{\aleph}(A)$ has to be
accepted if the left part in (\ref{t1}) or (\ref{cr})
  is less than $- \log (\alpha \omega_i).$
Obviously, $- \log (\alpha \omega_i)$ goes to infinity if $i$
increases. That is why there are many cases, where   it is enough
to check a 
 finite number of hypotheses
$H_0^{\aleph}(\Lambda_i)$.

\begin{theorem}\label{testinf} i) For each  $\alpha \in (0,1),$
sequence of partitions $\hat{\Lambda}$ and  a code $\varphi$, the
Type I error of the described test $\textbf{T}_{
\alpha,\varphi}^{\aleph}(\hat{\Lambda})$ is not larger than
$\alpha $, and ii) if, in addition, $\varphi$ is a universal code
and $\hat{\Lambda}$ discriminates between $H_0^\aleph(A),
H_1(A)^\aleph,$ then the Type II error of the test $\textbf{T}_{
\alpha,\varphi}^{\aleph}(\hat{\Lambda})$ goes to 0, when the
sample size tends to infinity.
\end{theorem}

\section{Conclusion}

Time series  is a popular model of real stochastic processes which
has a lot of applications in industry, economy, meteorology and
many other fields. Despite this, there are many practically
important problems of statistical analysis of time series which
are still open. Among them we can name the problem of estimation
of the limiting probabilities and densities,  on-line prediction,
regression, classification  and some problems of hypothesis
testing (goodness-of-fit testing and testing of serial
independence).  This chapter describes a new approach to all the
problems mentioned above, which, on the one hand, gives a
possibility to solve the problems in the framework of the
classical mathematical statistics and, on the other hand, allows
to apply methods of real data compression  to solve these problems
in practise. Such applications to randomness testing \cite{RM} and
prediction of currency exchange rates \cite{RM1}  showed high
efficiency, that is why the suggested methods look very promising
for practical applications.  Of course,  problems like prediction
of price of oil, gold, etc. and testing of different random number
generators can be used as case studies for students.

\section{Appendix}

\begin{proof}[Claim \ref{Lap}] We employ
the general inequality
$$
D(\mu\|\eta)\leq   \log e \,\,(-1+\sum_{a\in A}\mu(a)^2/ \eta(a)\,
),
$$
 valid for any distributions $\mu$ and $\eta$ over $A$  (follows from
the elementary inequality for natural logarithm  $\ln x \leq
x-1)$, and find:
 $$
\rho^t(P\|L_0) = \sum_{x_1\cdots x_t\in A^t}P(x_1\cdots
x_t)\,\,\sum_{a\in A}P(a|x_1\cdots x_t)\log\frac{P(a|x_1\cdots
x_t)} {\gamma(a|x_1\cdots x_t)}$$
$$= \log e\, (\sum_{x_1\cdots x_t\in
A^t}P(x_1\cdots x_t)\,\,\sum_{a\in A}P(a|x_1\cdots x_t)\ln
\frac{P(a|x_1\cdots x_t)} {\gamma(a|x_1\cdots x_t)}) $$
$$\leq \log e \, (-1+\sum_{x_1\cdots x_t\in A^t}P(x_1\cdots x_t) \sum_{a\in A}\frac
{P(a)^2(t+ |A|)}{\nu_{x_1\cdots x_t}(a)+1}$$ Applying the
well-known Bernoulli formula, we obtain
$$ \rho^t(P\|L_0) = \log e \, ( -1+\sum_{a\in A}\sum_{i=0}^t\frac
{P(a)^2(t+|A|)}{i+1} \, \bigl(\begin {array}{c} t\\
i\end{array}\bigr) P(a)^i(1-P(a))^{t-i} )
$$ $$ = \log e \, ( -1+\frac{t+|A|}{t+1}\sum_{a\in A} P(a) \sum_{i=0}^t
\bigl( \begin{array}{c} t+1\\
i+1\end{array}\bigr) P(a)^{i+1}(1-P(a))^{t-i} ) $$
$$\leq  \log e \, ( -1+\frac{t+|A|}{t+1}\sum_{a\in A}P(a) \sum_{j=0}^{t+1}
\bigl( \begin{array}{c} t+1 \\j\end{array} \bigr)
P(a)^j(1-P(a))^{t+1-j} ) . $$ Again, using the Bernoulli formula,
we finish the proof
$$ \rho^t(P\|L_0) = \log e \, \frac{|A|-1}{t+1}.
$$
The second statement of the claim follows from the well-known
asymptotic equality $$ 1 + 1/2 + 1/3 + ... + 1/t = \ln t + O(1),
$$ the obvious presentation
$$
\bar{\rho}^t(P\|L_0) = t^{-1} (\rho^0(P\|L_0) + \rho^1 (P\|L_0)+
... + \rho^{t-1}(P\|L_0))
$$ and (\ref{rL}).
\end{proof}

\begin{proof}[Claim \ref{c2}]
The first equality follows from the definition (\ref{RR}), whereas
the second from the definition (\ref{h0}). From (\ref{Kp1}) we
obtain:
$$ - \log
K_0(x_1 ... x_t) = - \log (
\frac{\Gamma(|A|/2)}{\Gamma(1/2)^{|A|}}\: \frac{\prod_{a \in A}
\Gamma(\nu^t(a)+ 1/2 )}{\Gamma((t+ |A|/2 )} )
$$
$$ = c_1 + c_2 |A| + \log \Gamma (t+ |A|/2) - \sum_{a \in A}
\Gamma(\nu^t(a) + 1/2) ,
$$ where $c_1 , c_2$ are constants.  Now we use the well known Stirling formula
$$ \ln \Gamma(s) = \ln \sqrt{2 \pi} + (s- 1/2)\ln s   - s +
\theta/ 12, $$ where $\theta  \in (0,1)$  \cite{Kn} . Using this
formula we rewrite the previous equality as
$$ - \log
K_0 (x_1 ... x_t) = - \sum_{a \in A} \nu^t(a) \log (\nu^t(a)/t) +
(|A|-1)\log t /2 + \bar{c}_1 + \bar{c}_2 |A| ,$$ where $\bar{c}_1
, \bar{c}_2$ are constants.
Hence,
$$ \sum_{x_1 \ldots x_t \in A^t}P(x_1 \ldots x_t)( - \log (K_0 (x_1 ...
x_t) )) $$ $$ \leq t ( \sum_{x_1 \ldots x_t \in A^t}P(x_1 \ldots
x_t) (\, - \sum_{a \in A} \nu^t(a) \log (\nu^t(a)/t) )+
(|A|-1)\log t /2 +  c |A| .$$   Applying  the well known Jensen
inequality for the concave function $ - x \log x $ we obtain the
following inequality:
$$ \sum_{x_1 \ldots x_t \in A^t}P(x_1 \ldots x_t)( - \log (K_0 (x_1 ...
x_t) ) \leq  $$ $$   - t (\sum_{x_1 \ldots x_t \in A^t} P(x_1
\ldots x_t) ((\nu^t(a) / t)) $$ $$ \log \sum_{x_1 \ldots x_t \in
A^t}P(x_1 \ldots x_t) (\nu^t(a) / t) + (|A|-1)\log t /2 +  c |A| .
$$ The source $P$ is i.i.d.,  that is why the average
frequency $$\sum_{x_1 \ldots x_t \in A^t}P(x_1 \ldots x_t)
\nu^t(a)$$ is equal to $P(a)$ for any $a \in A $ and we obtain
from two last formulas the following inequality:
\begin{equation}\label{1} \sum_{x_1 \ldots x_t \in A^t}P(x_1 \ldots x_t)( - \log (K_0(x_1 ...
x_t) ) $$ $$  \leq t (- \sum_{a \in A} P(a) \log P(a)) +
(|A|-1)\log t /2 + c |A|
\end{equation}
On the other hand,
\begin{equation}\label{2} \sum_{x_1 \ldots x_t \in A^t}P(x_1 \ldots x_t)(  \log P(x_1 \ldots
x_t)) = \sum_{x_1 \ldots x_t \in A^t}P(x_1 \ldots x_t)
\sum_{i=1}^t \log P(x_i) $$ $$ = t (\sum_{a \in A} P(a) \log
P(a)).
\end{equation} From (\ref{1}) and (\ref{2}) we can see that
$$ t^{-1} \sum_{x_1 \ldots x_t \in A^t}P(x_1 \ldots x_t) \log \frac{P(x_1 \ldots
x_t)}{(K_0(x_1 ... x_t) } \leq ((|A|-1)\log t /2 + c)/t . $$
\end{proof}
\begin{proof}[Claim \ref{kmc}]
First we consider the case where $m=0.$ The proof for this case is
very close to the proof of the previous claim. Namely, from
(\ref{Kp1}) we obtain:
$$ - \log
K_0(x_1 ... x_t) = - \log (
\frac{\Gamma(|A|/2)}{\Gamma(1/2)^{|A|}}\: \frac{\prod_{a \in A}
\Gamma(\nu^t(a)+ 1/2 )}{\Gamma((t+ |A|/2 )} )
$$
$$ = c_1 + c_2 |A| + \log \Gamma (t+ |A|/2) - \sum_{a \in A}
\Gamma(\nu^t(a) + 1/2) ,
$$ where $c_1 , c_2$ are constants.  Now we use the well known Stirling formula
$$ \ln \Gamma(s) = \ln \sqrt{2 \pi} + (s- 1/2)\ln s   - s +
\theta/ 12, $$ where $\theta  \in (0,1)$   \cite{Kn} . Using this
formula we rewrite the previous equality as
$$ - \log
K_0 (x_1 ... x_t) = - \sum_{a \in A} \nu^t(a) \log (\nu^t(a)/t) +
(|A|-1)\log t /2 + \bar{c}_1 + \bar{c}_2 |A| ,$$ where $\bar{c}_1
, \bar{c}_2$ are constants. Having taken into account the
definition of the empirical entropy (\ref{He}), we obtain
$$ - \log
K_0(x_1 ... x_t) \leq t h^*_0( x_1 \ldots x_t) + (|A|-1)\log t /2
+ c |A| .$$ Hence,
$$ \sum_{x_1 \ldots x_t \in A^t}P(x_1 \ldots x_t)( - \log (K_0 (x_1 ...
x_t) )) $$ $$ \leq t ( \sum_{x_1 \ldots x_t \in A^t}P(x_1 \ldots
x_t) h^*_0( x_1 \ldots x_t) + (|A|-1)\log t /2 +  c |A| .$$ Having
taken into account the definition of the empirical entropy
(\ref{He}), we apply the well known Jensen inequality for the
concave function $ - x \log x $ and obtain the following
inequality:
$$ \sum_{x_1 \ldots x_t \in A^t}P(x_1 \ldots x_t)( - \log (K_0 (x_1 ...
x_t) ) \leq +  c |A|   \:- $$ $$ t (\sum_{x_1 \ldots x_t \in
A^t}P(x_1 \ldots x_t) ((\nu^t(a) / t)) \log \sum_{x_1 \ldots x_t
\in A^t}P(x_1 \ldots x_t) (\nu^t(a) / t) + (|A|-1)\log t /2  .
$$ $P$ is stationary and ergodic,  that is why the average
frequency $$\sum_{x_1 \ldots x_t \in A^t}P(x_1 \ldots x_t)
\nu^t(a)$$ is equal to $P(a)$ for any $a \in A $ and we obtain
from two last formulas the following inequality:
$$ \sum_{x_1 \ldots x_t \in A^t}P(x_1 \ldots x_t)( - \log (K_0(x_1 ...
x_t) )  \leq t\: h_0(P) + (|A|-1)\log t /2 +  c |A|, $$ where
$h_0(P) $ is the first order Shannon entropy, see  (\ref{h0}).

 We have
seen that any source from $M_m(A)$ can be presented as a "sum" of
$|A|^m$ i.i.d. sources. From this we can easily see that the error
of a predictor for the source from $M_m(A)$ can be upper bounded
by the error of i.i.d. source multiplied by $|A|^m$. In
particular, we obtain from the last inequality and the definition
of the Shannon entropy (\ref{moe}) the  upper bound (\ref{c3}).
\end{proof}
\begin{proof}[Theorem~\ref{TR}]
We can see from the definition (\ref{R}) of
 $R$  and the Claim~\ref{km}  that the average error
  is upper bounded as
follows: $$ - \;  t^{-1} \sum_{x_1 ... x_t \in A^t} P(x_1 ... x_t)
\log ( R (x_1 ... x_t)) - h_k(P)$$ $$ \leq  (|A|^k (|A| - 1) \log
t + \log (1/ \omega_i) + C)/ (2 t) ,
$$ for any $k= 0, 1, 2, ...$. Taking into account that for any $P \in M_\infty(A)$
$\lim_{k\rightarrow\infty}h_k(P) = h_\infty(P),$ we can see that
$$ (\lim_{t\rightarrow\infty}t^{-1} \sum_{x_1 ... x_t \in A^t} P(x_1 ... x_t)
\log ( R (x_1 ... x_t)) - h_\infty(P)) = 0.$$ The second statement
of the theorem is proven. The first  one can be easily derived
from the ergodicity of $P$  \cite{Billingsley, Ga} .
\end{proof}
\begin{proof}[Theorem~\ref{KorotkoObozvatEtuTeoremu} ] The proof is based on the
 Shannon-MacMillan-Breiman theorem which states that
 for any stationary and ergodic source $P$
  $$
\lim_{t\rightarrow\infty} -  \log P(x_1 \ldots x_t) /t =
h_\infty(P)
$$
 with probability 1 \cite{Billingsley,Ga} .
From this equality and  (\ref{U1}) we obtain the statement i). The
second statement follows from the definition of the Shannon
entropy (\ref{hlim}) and (\ref{U2}).
\end{proof}
\begin{proof}[Theorem~\ref{t4}]   i) immediately
follows from the second statement of the
theorem~\ref{KorotkoObozvatEtuTeoremu} and properties of $\log$.
The statement ii) can be proven as follows:

$$    \,   \lim_{t\rightarrow\infty}
E(  \frac {1}{t}   \, \,   \sum_{i=0}^{t-1} ( P(x_{i+1}|x_1 \ldots
x_i)  - \mu_U(x_{i+1}|x_1 \ldots x_i) )^2)  =$$
$$\lim_{t\rightarrow\infty}   \frac {1}{t}   \, \,
\sum_{i=0}^{t-1} \sum_{x_1 \ldots x_i \in A^i}P(x_1 \ldots x_i) (
\sum_{a \in A} |P(a|x_1 \ldots x_i)  - \mu_U(a|x_1 \ldots x_i)|
)^2 \leq  $$
$$\lim_{t\rightarrow\infty}   \frac {const}{t}   \, \,
\sum_{i=0}^{t-1}\sum_{x_1 \ldots x_i \in A^i}P(x_1 \ldots x_i)
 \sum_{a \in A} P(a|x_1 \ldots x_i)  \log \frac{P(a|x_1 \ldots x_i)}{
\mu_U(a|x_1 \ldots x_i) }   = $$
$$ \lim_{t\rightarrow\infty}(  \frac {const}{t}   \, \,   \sum_{ x_1 \ldots x_t  \in A^t}
P(x_1 \ldots x_t)  \log (P(x_1 \ldots x_t)  / \mu(x_1 \ldots x_t)
) ). $$ Here the first inequality is obvious, the second follows
from the Pinsker's inequality (\ref{pi}), the others from
properties of expectation  and $\log .$ iii) can be derived from
ii) and the Jensen inequality for the function $x^2. $
\end{proof}
\begin{proof}[Theorem~\ref{t5}] The following inequality follows from the nonnegativity of
the KL divergency (see (\ref{pi})), whereas the equality is
obvious.
$$ E( \log  \frac{ P(x_1|y_1) } { \mu_U(x_1|y_1) }) +
E(\log  \frac{ P(x_2|(x_1,y_1),y_2) } { \mu_U(x_2|(x_1,y_1),y_2)
}) +  \ldots \leq E( \log  \frac{ P(y_1) } { \mu_U(y_1) })$$
$$ + E(
\log  \frac{ P(x_1|y_1) } { \mu_U(x_1|y_1) })+ E(\log  \frac{
P(y_2|(x_1,y_1) } { \mu_U(y_2|(x_1,y_1) }) + E(\log \frac{
P(x_2|(x_1,y_1),y_2) } { \mu_U(x_2|(x_1,y_1),y_2) }) + \ldots  $$
$$ = E( \log  \frac{ P(x_1,y_1) } { \mu_U(x_1,y_1) }) + E(\log
\frac{ P((x_2,y_2)|(x_1,y_1)) } { \mu_U((x_2,y_2)|(x_1,y_1)) }) +
... .$$ Now we can apply the first statement of the previous
theorem to the last sum as follows:
$$ \lim_{t\rightarrow\infty}   \frac {1}{t} E( \log  \frac{ P(x_1,y_1) } { \mu_U(x_1,y_1) })
+ E(\log  \frac{ P((x_2,y_2)|(x_1,y_1)) } {
\mu_U((x_2,y_2)|(x_1,y_1)) }) + ... $$
$$ E(\log  \frac{ P((x_t,y_t)|(x_1,y_1)\ldots (x_{t-1},y_{t-1})) }
{ \mu_U((x_t,y_t)|(x_1,y_1)\ldots (x_{t-1},y_{t-1})) }) = 0.$$
From this equality and the last inequality we obtain the proof of
i). The proof of the second statement can be obtained from the
similar representation for ii) and the second statement of the
theorem 4. iii) can be derived from ii) and the Jensen inequality
for the function $x^2.$ \end{proof}

\begin{proof}[Lemma~\ref{Le}] . First we show that for any source
$\theta^* \in M_0(A)$ and any words $x^1 = x^1_1 ... x^1_{t_1},$
$... ,$ $x^r = x^r_1 ... x^r_{t_r},$
$$\theta^* (x^1 \diamond ... \diamond x^r) = \prod_{a \in A}
(\theta^*(a))^{\nu_{x^1 \diamond ... \diamond x^r}(a)} $$
\begin{equation}\label{ta} \leq \prod_{a \in A} (
\nu_{x^1 \diamond ... \diamond x^r}(a)/t)^{\nu_{x^1 \diamond ...
\diamond x^r}(a)} ,
\end{equation} where $t = \sum_{i=1}^r t_i .$
Here the equality holds, because $\theta^* \in M_0(A)$ . The
inequality follows from  Claim 1. Indeed, if $ p(a) = \nu_{x^1
\diamond ... \diamond x^r}(a)/t$ and $ q(a) = \theta^* (a),$ then
$$ \sum_{a \in A} \frac{\nu_{x^1 \diamond ... \diamond x^r}(a)}{t}
\log \frac{(\nu_{x^1 \diamond ... \diamond x^r}(a)/t)}{\theta^*(a)
}\geq 0. $$ From the latter inequality we obtain (\ref{ta}).
Taking into account the definition (\ref{He1}) and (\ref{ta}), we
can see that the statement of Lemma is true for this particular
case.

For any $\theta \in M_m(A)$ and $x = x_1 \ldots x_s, \,s > m,$ we
present $\theta (x_1 \ldots x_s)$ as $$\theta (x_1 \ldots x_s)=
\theta(x_1 \ldots x_m) \prod_{u \in A^m } \prod_{a \in A} \theta
(a/u)^{\nu_x(ua)}\:,$$ where $\theta(x_1 \ldots x_m)$
 is
the limiting probability of the word $x_1 \ldots x_m .$ Hence,
$\theta (x_1 \ldots x_s) \leq \prod_{u \in A^m } \prod_{a \in A}
\theta (a/u)^{\nu_x(ua)}\:.$ Taking into account the inequality
(\ref{ta}), we obtain $\prod_{a \in A} \theta (a/u)^{\nu_x(ua)}
\leq \prod_{a \in A} ( \nu_x(ua)/\bar{\nu}_x(u))^{\nu_x(ua)}$
 for any word $u$. Hence, $$ \theta (x_1 \ldots x_s) \leq \prod_{u \in A^m }
 \prod_{a \in A} \theta (a/u)^{\nu_x(ua)} $$ $$
\leq  \prod_{u \in A^m } \prod_{a \in A} (
\nu_x(ua)/\bar{\nu}_x(u))^{\nu_x(ua)}.$$  If we apply those
inequalities to $\theta(x^1 \diamond ... \diamond x^r),$ we
immediately obtain the following inequalities
$$ \theta(x^1 \diamond ... \diamond x^r)  \leq \prod_{u \in A^m }
 \prod_{a \in A} \theta (a/u)^{\nu_{x^1 \diamond ... \diamond x^r}(ua)}
\leq $$ $$ \prod_{u \in A^m } \prod_{a \in A} ( \nu_{x^1 \diamond
... \diamond x^r}(ua)/\bar{\nu}_{x^1 \diamond ... \diamond
x^r}(u))^{\nu_{x^1 \diamond ... \diamond x^r}(ua)}.$$ Now the
statement of the Lemma follows from the definition  (\ref{He1}).
\end{proof}

\begin{proof}[Theorem~\ref{test1}]  Let $C_\alpha$ be a critical set of
the test $T_{\varphi}^{\,id}(A, \alpha)$, i.e., by definition, $
C_\alpha = \{ u: u \in A^t \,\, \:\& \: - \log \pi(u) - |\varphi
(u) | > - \log \alpha  \}. $ Let $\mu_\varphi$ be a measure for
which the claim 2 is true. We define an auxiliary set $
\hat{C}_\alpha  $ $ = \{ u:  - \log \pi(u) - (- \log
\mu_\varphi(u) ) $ $ > - \log \alpha  \}. $ We have $ 1 \geq $ $
\sum_{u \in \hat{C}_\alpha} \mu_\varphi(u) $ $  \geq \sum_{u \in
\hat{C}_\alpha} \pi(u) / \alpha $ $ = (1/ \alpha)
\pi(\hat{C}_\alpha) .$ (Here the second inequality follows from
the definition of $\hat{C}_\alpha,$ whereas all others are
obvious.) So, we obtain that $\pi(\hat{C}_\alpha) \leq \alpha .$
From  definitions of $C_\alpha, \hat{C}_\alpha $ and (\ref{kra})
we immediately obtain that $\hat{C}_\alpha \supset C_\alpha.$
Thus, $\pi(C_\alpha) \leq \alpha .$ By definition, $\pi(C_\alpha)$
is the value of the Type I error. The first statement of the
theorem  is proven.

Let us prove the second statement of the theorem. Suppose that the
hypothesis $H_1^{id}(A)$ is true. That is, the sequence $x_1
\ldots x_t$ is generated by some stationary and ergodic source
$\tau$ and $\tau \neq \pi .$ Our strategy is to show that
\begin{equation}\label{inf}    \lim _{t\rightarrow\infty}- \log \pi(x_1 \ldots
x_t) - |\varphi(x_1 \ldots x_t)| = \infty
\end{equation}
 with probability 1 (according to the
measure $\tau$). First we represent (\ref{inf}) as
$$
 - \log \pi(x_1
\ldots x_t) - |\varphi(x_1 \ldots x_t)| $$ $$  = t ( \frac{1}{t}
\log \frac{\tau(x_1 \ldots x_t)}{\pi(x_1 \ldots x_t)}   +
\frac{1}{t}( - \log \tau(x_1 \ldots x_t)  - |\varphi(x_1 \ldots
x_t)| ) ). $$
 From
this equality and the property of a universal code (\ref{U1}) we
obtain
\begin{equation}\label{inf2}  - \log \pi(x_1 \ldots
x_t) - |\varphi(x_1 \ldots x_t)| = t\, ( \frac{1}{t} \log
\frac{\tau(x_1 \ldots x_t)}{\pi(x_1 \ldots x_t)}  + o(1)).
\end{equation}
 From  (\ref{U1}) and (\ref{hlim}) we can see that
\begin{equation}\label{inf3} \lim_{t\rightarrow\infty} -  \log \tau(x_1 \ldots x_t)
/t   \leq h_k( \tau)
\end{equation} for any $k \geq 0$ (with probability 1).
It is supposed that the process $\pi$ has a finite memory, i.e.
belongs to $M_s(A)$ for some $s$. Having taken into account the
definition of $M_s(A)$ (\ref{ma}), we obtain the following
representation: $$ - \log \pi(x_1 \ldots x_t)/t  =   - t^{-1}
\sum_{i=1}^t \log \pi(x_i/x_1 \ldots x_{i-1}) $$ $$  =  - t^{-1}
(\sum_{i=1}^k \log \pi(x_i/x_1 \ldots x_{i-1})  + \sum_{i=k+1}^t
\log \pi(x_i/x_{i-k} \ldots x_{i-1})) $$ for any $k \geq s.$
According to the ergodic theorem there exists a limit $$
\lim_{t\rightarrow\infty} t^{-1} \sum_{i=k+1}^t \log
\pi(x_i/x_{i-k} \ldots x_{i-1}),$$
 which  is equal to $ h_k(\tau)$
\cite{Billingsley, Ga} . So, from the two last equalities we can
see that $$  \lim_{t\rightarrow\infty} (- \log \pi(x_1 \ldots
x_t))/t = - \sum_{v \in A^k} \tau(v) \sum_{a \in A} \tau(a/v) \log
\pi(a/v).$$ Taking into account this equality, (\ref{inf3}) and
(\ref{inf2}), we can see that $$- \log \pi(x_1 \ldots x_t) -
|\varphi(x_1 \ldots x_t)|   \geq  t \,(\sum_{v \in A^k} \tau(v)
\sum_{a \in A} \tau(a/v) \log (\tau(a/v) / \pi(a/v) )) + o(t) $$
for any $k \geq s.$
   From this
inequality and Claim~\ref{Lap}  we can obtain that  $$\:- \log
\pi(x_1 \ldots x_t) - |\varphi(x_1 \ldots x_t)| \geq c\: t +
o(t)$$, where $c$ is a positive constant, $t\rightarrow\infty.$
Hence, (\ref{inf}) is true and the theorem is proven.
\end{proof}

\begin{proof}[Theorem~\ref{test2} ] Let us denote the critical set of the
test $T_{\, \varphi}^{\,SI}(A,\alpha)$  as $C_\alpha,$ i.e., by
definition, $ C_\alpha = \{ x_1 \ldots x_t :\; (t - m)\: h^*_m(x_1
\ldots x_t) - |\varphi(x_1 ... x_t)| )
>
\log (1 / \alpha)   \} . $ From  Claim~\ref{c2}  we can see that
there exists such a measure $\mu_\varphi $ that $ - \log
\mu_\varphi(x_1 ... x_t)$ $ \leq |\varphi(x_1 ... x_t)| \,. $ We
also define
\begin{equation}\label{C} \hat{C}_\alpha = \{ x_1 \ldots x_t :\;
(t - m)\: h^*_m(x_1 \ldots x_t) - (- \log \mu_\varphi(x_1 ...
x_t)) \,)
  >
\log (1 / \alpha)   \} . \end{equation} Obviously, $\hat{C}_\alpha
\supset C_\alpha .$ Let $\theta$ be any source from $M_m(A).$ The
following chain of equalities and inequalities is true: $$ 1 \geq
\mu_\varphi(\hat{C}_\alpha) = \sum_{x_1 \ldots x_t \in
\hat{C}_\alpha} \mu_\varphi(x_1 \ldots x_t) $$ $$ \geq \alpha^{-1}
\sum_{x_1 \ldots x_t \in \hat{C}_\alpha} 2^{(t-m) h_m^*(x_1 \ldots
x_t)} \geq \alpha^{-1} \sum_{x_1 \ldots x_t \in \hat{C}_\alpha}
\theta(x_1 \ldots x_t) = \theta(\hat{C}_\alpha).$$
  (Here both equalities and the first inequality are
 obvious, the second   and the third inequalities follow from
(\ref{C}) and the Lemma, correspondingly.) So, we obtain that
$\theta(\hat{C}_\alpha) \leq \alpha $ for any source $\theta \in
M_m(A).$ Taking into account that $\hat{C}_\alpha \supset
C_\alpha,$ where $C_\alpha$  is the critical set of the test, we
can see that the probability of the  Type I error is not greater
than $\alpha.$ The first statement of the theorem is proven.

The proof of the second  statement  will  be based on some results
of Information Theory. We obtain from (\ref{U1}) that for any
stationary and ergodic $p$
\begin{equation}\label{unh}
\lim_{t\rightarrow\infty}  t^{-1} |\varphi(x_1 ... x_t)| =
h_\infty (p)
\end{equation} with probability 1.
    It can be seen from (\ref{He}) that $h^*_m$
 is an estimate for the $m-$order Shannon
entropy  (\ref{moe}). Applying the ergodic theorem we obtain $
\:\lim_{t\rightarrow\infty} h^*_m (x_1 \ldots x_t ) = h_m(p)$ with
probability 1  \cite{Billingsley, Ga} . It is known in Information
Theory that $h_m(\varrho) - h_\infty(\varrho)
> 0, $ if $\varrho$ belongs to $M_\infty(A)\: \backslash \:M_m(A)$
  \cite{Billingsley, Ga} . It is supposed that $H_1^{SI}$ is
true, i.e. the considered process belongs to $M_\infty(A)\:
\backslash\: M_m(A).$ So, from (\ref{unh}) and  the last equality
we obtain that $\lim_{t\rightarrow\infty} ( (t - m) \,h^*_m (x_1
\ldots x_t )  - |\varphi(x_1 ... x_t)|) = \infty .$ This proves
the second statement of the theorem.
\end{proof}

\begin{proof}[Theorem~\ref{t6}]
 First we prove that with probability 1 there exists the following limit
$ \lim_{t\rightarrow\infty}  \frac{1}{t}  \log (p(x_1 \ldots x_t)
/ r_U(x_1 \ldots x_t)) \, $ and this limit is finite and
nonnegative.
Let $A_n=\{x_1,\dots,x_n: p(x_1,\dots,x_n)\ne0\}$. Define
\begin{equation}\label{z} z_n(x_1 \ldots x_n)
= r_U(x_1 \ldots x_n) /p(x_1 \ldots x_n)
\end{equation}
for $(x_1,\dots,x_n)\in A$ and $z_n=0$ elsewhere.

Since 
$$
 E_P(z_n| x_1,\dots,x_{n-1})= E\left(\frac{r_U(x_1 \ldots x_n)}{p(x_1 \ldots x_n)} \middle| x_1,\dots,x_{n-1}\right)
$$
$$=
\frac {r_U(x_1 \ldots x_{n-1})}{p(x_1 \ldots x_{n-1})}
E_P\left(\frac{r_U(x_n|x_1 \ldots x_{n-1})}{p(x_n| x_1 \ldots
x_{n-1})}\right)
$$
$$=
z_{n-1}  \int_A \frac{r_U(x_n|x_1 \ldots x_{n-1}) d P(x_n| x_1
\ldots x_{n-1})}{d P(x_n| x_1 \ldots x_{n-1})/ d M_n (x_n| x_1
\ldots x_{n-1})}
$$$$
= z_{n-1}  \int_A r_U(x_n|x_1 \ldots x_{n-1})d M_n (x_n| x_1
\ldots x_{n-1}) \le z_{n-1}
$$
the stochastic sequence $(z_n,B^n)$ is, by definition,  a
non-negative supermartingale with respect to $P$, with  $E(z_n)
\le1$,  \cite{Shir} . Hence, Doob's submartingale convergence
theorem implies that  the limit $z_n$ exists and is finite with
$P-$probability 1 (see \cite[Theorem 7.4.1]{Shir}). Since all
terms are nonnegative so is the limit. Using the
definition~(\ref{z})  with $P$-probability~1 we have
$$ \lim_{n\rightarrow\infty}
 p(x_1 \ldots x_n)/ r_U(x_1 \ldots x_n)  >   0 ,  $$
 $$ \lim_{n\rightarrow\infty}
\log ( p(x_1 \ldots x_n)/ r_U(x_1 \ldots x_n) ) >   - \infty  $$
and
\begin{equation}\label{th10}  \lim_{n\rightarrow\infty} n^{-1}\, \log ( p(x_1 \ldots x_n)/ r_U(x_1 \ldots
x_n) ) \geq   0. \end{equation} 
 Now we note that for any integer $s$ the following obvious equality is true:
  $r_U(x_1 \ldots x_t) = \omega_s \mu_U(x_1^{[s]} \ldots x_t^{[s]}) /
M_t(x_1^{[s]} \ldots x_t^{[s]})  \:(1+ \delta)$ for some $\delta
> 0.$ From this equality, (\ref{UC}) and (\ref{den}) we immediately obtain that
a.s.
\begin{equation}\label{th1}
\lim_{t\rightarrow\infty}  \frac{1}{t}  \log \frac{p(x_1 \ldots
x_t)}{r_U(x_1 \ldots x_t)} \, \:\leq  \lim_{t\rightarrow\infty}
\frac{- \log \omega_t}{t}   $$ $$ +  \lim_{t\rightarrow\infty}
\frac{1}{t}
  \log \frac{p(x_1 \ldots x_t)}{\mu_U(x_1^{[s]} \ldots x_t^{[s]}) / M_t(x_1^{[s]} \ldots
x_t^{[s]})  } \, $$ $$ \leq  \lim_{t\rightarrow\infty} \frac{1}{t}
  \log \frac{p(x_1 \ldots x_t)}{2^{-
|U(x_1^{[s]} \ldots x_t^{[s]})| } / M_t(x_1^{[s]} \ldots
x_t^{[s]})  } . \end{equation} The right part  can be presented as
follows:
 \begin{equation}\label{th2} \lim_{t\rightarrow\infty} \frac{1}{t} \log \frac{p(x_1 \ldots x_t)}{2^{-
|U(x_1^{[s]} \ldots x_t^{[s]})| } / M_t(x_1^{[s]} \ldots
x_t^{[s]})  } $$ $$ = \lim_{t\rightarrow\infty} \frac{1}{t}
  \log \frac{p^s (x_1 \ldots x_t)\: M_t(x_1^{[s]} \ldots
x_t^{[s]}) }{2^{- |U(x_1^{[s]} \ldots x_t^{[s]})| } }
\end{equation}
$$ + \lim_{t\rightarrow\infty} \frac{1}{t} \log \frac{p(x_1 \ldots x_t)}{p^s (x_1 \ldots
x_t)}.
$$
Having taken into account  that $U$ is a universal code, 
(\ref{ds}) and the theorem~\ref{KorotkoObozvatEtuTeoremu}, we can
see that the  first term is equal to zero. From (\ref{ba})
 and (\ref{Ba2})  we can see that a.s. the second term is
equal to $\tilde{h}_s - \tilde{h}.$ This equality is valid for any
integer $s$ and, according to (\ref{hh}),  the second term equals
zero, too, and we obtain that
$$\lim_{t\rightarrow\infty}  \frac{1}{t}  \log \frac{p(x_1 \ldots
x_t)}{r_U(x_1 \ldots x_t)} \, \:\leq 0 . $$
 Having taken into account (\ref{th10}),
we can see that the first statement is proven.

 From     (\ref{th1})  and (\ref{th2})
we can can see that
\begin{equation}\label{th6}
 E \,  \log \frac{p(x_1 \ldots
x_t)}{r_U(x_1 \ldots x_t)} \leq  E\,
  \log \frac{p^s_t (x_1, \ldots, x_t)\:M_t(x_1^{[s]} \ldots
x_t^{[s]}) }{2^{- |U(x_1^{[s]} \ldots x_t^{[s]})| } } $$ $$  + E\:
 \log \frac{p(x_1 \ldots x_t)}{p^s (x_1, \ldots, x_t)}
.  \end{equation} The first term is the average redundancy of the
universal code for a finite- alphabet source,  hence, according to
the theorem~\ref{KorotkoObozvatEtuTeoremu}, it tends to 0. The
second term tends to      $\tilde{h}_s - \tilde{h}$ for any $s$
and from (\ref{hh}) we can see that it is equals to zero. The
second statement is proven.
\end{proof}
\begin{proof}[Theorem~\ref{t7}]  Obviously,
 \begin{equation}\label{tha}
 E( \frac{1}{t} \sum_{m=0}^{t-1} ( P(x_{m+1} \in B_{m+1}  | x_1 ... x_m) -
 R_U(x_{m+1} \in B_{m+1}  | x_1 ... x_m) )^2 ) \leq
\end{equation}
$$ \frac{1}{t} \sum_{m=0}^{t-1} E ( | P(x_{m+1} \in B_{m+1}  | x_1 ... x_m) -
 R_U(x_{m+1} \in B_{m+1}  | x_1 ... x_m) |
+ $$
$$  | P(x_{m+1} \in \bar{ B}_{m+1}  | x_1 ... x_m) -
 R_U(x_{m+1} \in\bar{ B}_{m+1}  | x_1 ... x_m) | )^2 . $$
From the Pinsker inequality     (\ref{pi}) and    convexity of the
KL divergence   (\ref{r0}) we obtain the following inequalities
\begin{equation}\label{thb}
 \frac{1}{t} \sum_{m=0}^{t-1} E ( | P(x_{m+1} \in B_{m+1}  | x_1 ... x_m) -
 R_U(x_{m+1} \in B_{m+1}  | x_1 ... x_m) |
+
\end{equation}
$$  | P(x_{m+1} \in \bar{ B}_{m+1}  | x_1 ... x_m) -
 R_U(x_{m+1} \in\bar{ B}_{m+1}  | x_1 ... x_m) | )^2 \leq $$
$$ \frac{ const}{t} \sum_{m=0}^{t-1}
E( ( \log \frac{ P(x_{m+1} \in B_{m+1}  | x_1 ... x_m)}{
R_U(x_{m+1} \in B_{m+1}  | x_1 ... x_m)} +
 \log  \frac{  P(x_{m+1} \in \bar{ B}_{m+1}  | x_1 ... x_m) }  { R_U(x_{m+1} \in\bar{ B}_{m+1}  | x_1 ... x_m) } )
  \leq $$
$$  \frac{const}{t} \sum_{m=0}^{t-1} ( \int   p( x_1 ... x_m) (\int p(  x_{m+1} | x_1 ... x_m) )
\log  \frac{p(  x_{m+1} | x_1 ... x_m) } {r_U(  x_{m+1} | x_1 ...
x_m) } d M ) d M_m ) .$$ Having taken into account that the last
term is equal to $ \frac{const}{t}  E( \log \frac{ p(x_1 ... x_t)
} { r_U( x_1 ... x_t) } ) , $  from     (\ref{tha}), (\ref{thb})
and (\ref{th}) we obtain   (\ref{thc}). ii) can be derived from i)
and the Jensen inequality for the function $x^2.$
\end{proof}
\begin{proof}[Theorem~\ref{t8}]   The last inequality of the following chain
follows from the Pinsker's one, whereas all others are obvious.
$$    ( \int  f(x)\, p(x  | x_1 ... x_m) \, d M_m  
-  \int  f(x)\, r_U(x  | x_1 ... x_m)\, d M_m  )^2    $$ $$ =  (
\int f(x)\, (p(x  | x_1 ... x_m)  - r_U(x  | x_1 ... x_m) ) \,d
M_m )^2
$$ $$  \leq \bar{M}^2  ( \int (p(x  | x_1 ... x_m)  - r_U(x | x_1
... x_m) )\, d M_m  )^2
$$   $$
\leq    \bar{M}^2  ( \int  |p(x  | x_1 ... x_m)  - r_U(x  | x_1
... x_m) |  d M_m  )^2  $$ $$\leq const  \,    \int  \, p(x | x_1
... x_m)\, \log \frac{ p(x  | x_1 ... x_m)} {r_U(x  | x_1 ... x_m)
}d M_m   .
$$
From these inequalities we obtain:
\begin{equation}\label{int1}
 E ( \sum_{m=0}^{t-1}   (  \int  f(x)\, p(x  | x_1 ... x_m) \,d M_m  - $$ $$ \int  f(x)\, r_U(x  | x_1 ... x_m)\,
d M_m  )^2 )\leq
\end{equation}
 $$  \sum_{m=0}^{t-1}
const  \,   E ( \int \,   p(x  | x_1 ... x_m) \,\log \frac{ p(x  |
x_1 ... x_m)} {r_U(x  | x_1 ... x_m) }   d M_{1/m}) .
$$
The last term can be presented as follows:
$$
\sum_{m=0}^{t-1}   E ( \int    p(x  | x_1 ... x_m)\, \log \frac{
p(x | x_1 ... x_m) } {r_U(x  | x_1 ... x_m) } d M_{1/m} )  = $$
$$            \sum_{m=0}^{t-1}   \int     p(x_1 ... x_m) $$ $$
 \int    p(x  | x_1 ... x_m) \log \frac{ p(x  | x_1 ... x_m)} {r_U(x  | x_1 ... x_m)}  d M_{1/m}  d M_m   $$
$$ =  \int  \,  p(x_1 ... x_t) \log (p(x_1 ... x_t)/r_U(x_1 ... x_t))  d M_t     . $$
From this equality,    (\ref{int1}) and   Corollary 1 we obtain
(\ref{int}). ii) can be derived from (\ref{int1}) and the Iensen
inequality for the function $x^2.$  \end{proof}

\begin{proof}[Theorem~\ref{testinf}]
 The following chain proves the first
statement of the theorem:
$$ P \{ H_0^{\aleph}(A)\quad is\; rejected \;/ H_0 \:is\: true\} =   P
\{\bigcup_{i=1}^\infty  \{H_0^\aleph(\Lambda_i)\quad is\;
rejected\;/ H_0 \:is\: true\} \:\}\,  $$
$$\leq \sum_{i=1}^\infty P
 \{ H_0^\aleph(\Lambda_i)\;/ H_0 \:is\: true\} \, \leq \sum_{i=1}^\infty (\alpha \omega_i) \, =
\alpha .
$$ (Here both inequalities follow from the description of the
test, whereas the last equality follows from (\ref{om}).)

The second statement also follows from the description of the
test. Indeed, let a sample is created by a source $\varrho,$ for
which $ H_1(A)^\aleph$ is true. It is supposed that the sequence
of partitions $\hat{\Lambda}$ discriminates between
$H_0^\aleph(A), H_1^\aleph(A).$ By definition, it means that there
exists $j$ for which $H_1^\aleph(\Lambda_j)$
 is true for the process $\varrho_{\Lambda_j}.$ It  immediately
 follows from Theorem~\ref{TR} - ~\ref{t4}
  that the Type II error
 of the test $T_{\,
\varphi}^{\,\aleph}(\Lambda_j,\alpha \omega_j)$ goes to 0, when
the sample size tends to infinity.
\end{proof}



\begin{thebibliography}{9}


\bibitem{Al}
P.~Algoet, Universal Schemes for Learning the Best Nonlinear
    Predictor Given the Infinite Past and Side Information,
  \emph{IEEE Trans. Inform. Theory,} \textbf{45},  1165-1185, (1999).


\bibitem{Bab}
 G.~J. Babu,   A.~Boyarsky, Y.~P. Chaubey,  P.~Gora, New statistical
method for filtering and entropy estimation of a chaotic map from
noisy data,   \emph{International Journal of Bifurcation and
Chaos,} \textbf{14} (11), 3989-3994, (2004).


\bibitem{Ba}
 A.R. Barron, The strong ergodic theorem for dencities: generalized
Shannon-McMillan-Breiman theorem,  \emph{The annals of
Probability},  \textbf{13} (4),  1292--1303, 1985.

\bibitem{BG}
L.Gy\"{o}rfi, I.P\'{a}li and E.C.  van der Meulen,  There is no
universal code for infinite alphabet,
  { \it IEEE Trans. Inform. Theory,}
 \textbf{40}, 267--271, 1994.

\bibitem{Billingsley}  P.~Billingsley,  \emph{Ergodic theory and
information}.
 (John Wiley \& Sons, 1965).


\bibitem{V1}
 R.~Cilibrasi and   P.~M.B. Vitanyi,
 Clustering by Compression,  \emph{IEEE Transactions on Information Theory}, \textbf{51} (4),
 (2005).

\bibitem{V2}
 R.~Cilibrasi, R.~de Wolf  and P.~M.B. Vitanyi,  Algorithmic Clustering of
 Music,
   \emph{Computer Music Journal}, \textbf{28} (4)  49--67, (2004).

\bibitem{CS}
I.~Csisz$\acute{a}$r and  P.~Shields, \textit{Notes on information
theory and statistics}. (Foundations and Trends in Communications
and Information Theory, 2004).

\bibitem{CS1}
I. ~Csisz$\acute{a}$r and  P. ~Shields,  The consistency of the
BIC Markov order estimation. \emph{Annals of Statistics,}
\textbf{6},
  1601-1619, 2000.


\bibitem{e} M.~Effros, K.~Visweswariah,  S.~R.Kulkarni and  S.~Verdu,
  Universal lossless source coding with the Burrows Wheeler
transform, \emph{IEEE Trans. Inform. Theory, } \textbf{45},
1315--1321,  (1999).

\bibitem{FE}
 W.~Feller,  \textit{An Introduction to Probabability Theory and Its
Applications}, vol.1. (John Wiley \& Sons, New York, 1970).

\bibitem{fin}
L. Finesso, C. Liu, and P. Narayan, The optimal error exponent for
Markov order estimation, \emph{IEEE Trans. Inf. Theory,}
\textbf{42}, (1996).

\bibitem{Fi}
B.~M. Fitingof,  Optimal encoding for unknown and changing
statistica of messages, \textit{Problems of Information
Transmission,} \textbf{2} (2),  3--11, (1966).

\bibitem{Ga}
 R.~G. Gallager,   \textit{Information Theory and Reliable Communication.}
(John Wiley \& Sons, New York, 1968).

\bibitem{gil}
E. N. Gilbert, Codes Based on Inaccurate Source Probabilities,
\emph{IEEE Trans. Inform. Theory, } \textbf{17}, (1971).

\bibitem{tcs}
N.Jevtic, A.Orlitsky and N.P.Santhanam. A lower bound on
compression of unknown alphabets. {\it Theoretical Computer
Science,} \textbf{332}, 293--311, 2004.


\bibitem{Ke} J.~L.  Kelly,  A new interpretation
of information rate, \emph{Bell System Tech. J.}, \textbf{35},
917--926, (1956).

\bibitem{Ki1}
J. ~Kieffer. A unified approach to weak universal source coding .
{\it IEEE Trans. Inform. Theory, } \textbf{24},  674--682, 1978.

\bibitem{Ki}
J.~Kieffer,    Prediction and Information Theory,
\textit{Preprint}, (available at
ftp://oz.ee.umn.edu/users/kieffer/papers/prediction.pdf/ ), 1998.


\bibitem{K-Y}
 J.~C. Kieffer and  En-Hui Yang, Grammar-based codes: a new class of
universal lossless source codes. \emph{IEEE Transactions on
Information Theory,} \textbf{46} (3), 737--754, (2000).


\bibitem{K}
A.~N. Kolmogorov,  Three approaches to the quantitative definition
of information, \textit{Problems Inform. Transmission},
\textbf{1}, 3--11, (1965).


\bibitem{Kn}
 D.~E.  Knuth { \it   The art of computer programming.} Vol.2.
(Addison Wesley, 1981).


\bibitem{Kr1}
 R.~ Krichevsky,  A relation between the plausibility of information
about a source and encoding redundancy, \emph{Problems Inform.
Transmission,} \textbf{4}(3),   48--57, (1968).



\bibitem{Kr}
R.~ Krichevsky,  {\it Universal Compression and Retrival}, (Kluver
Academic Publishers, 1993).

\bibitem{Ku}
S.~ Kullback,  \textit{Information Theory and Statistics.} (Wiley,
New York, 1959).

\bibitem{Ma}
U. ~Maurer,  Information-Theoretic Cryptography, In:
\emph{Advances in Cryptology - CRYPTO '99, Lecture Notes in
Computer Science,} Springer-Verlag, vol. 1666, pp. 47--64, (1999).

\bibitem{MM}
 D.~S. Modha and  E.~Masry,   Memory-universal prediction of stationary
random processes. \emph{IEEE Trans. Inform. Theory,}
\textbf{44}(1), 117--133, (1998).

\bibitem{No}
 A.~B. Nobel,   On optimal sequential prediction,  \textit{IEEE Trans.
Inform. Theory}, \textbf{49}(1),  83--98, (2003).

\bibitem{orl}
A. Orlitsky, N. P. Santhanam, and J. Zhang, Always Good Turing:
Asymptotically Optimal Probability Estimation, \textit{Science},
 \textbf{302}, (2003).


\bibitem{Rez}
Zh.~ Reznikova,  \textit{Animal Intelligence. From individual to
social cognition.} (CUP, 2007).

\bibitem{Ri1}
 J.~ Rissanen, Generalized Kraft inequality and arithmetic coding,
\emph{IBM J. Res. Dev.}, \textbf{20} (5),  198--203, (1976).

\bibitem{Ri}
 J.~ Rissanen,   Universal coding, information, prediction, and
estimation,   \emph{IEEE Trans. Inform. Theory, } 30(4), 629--636,
(1984).

\bibitem{rng}
 A.~ Rukhin and others. { \it A statistical test suite for random and
pseudorandom number generators for cryptographic applications. }
(NIST Special Publication 800-22 (with revision dated
May,15,2001)).
 http://csrc.nist.gov/rng/SP800-22b.pdf



\bibitem{Ry0}
  B.~Ya. Ryabko, Twice-universal coding, \textit{Problems of
Information Transmission,} \textbf{20}(3), 173--177, (1984).



\bibitem{Ry1}
 B. Ya. Ryabko,  Prediction of random sequences and universal coding.
\emph{Problems of Inform. Transmission}, \textbf{24}(2) 87-96,
(1988).

\bibitem{Ry2}
 B.~Ya. Ryabko,  A fast adaptive coding algorithm, \textit{Problems of
Inform. Transmission,} \textbf{26}(4), 305--317, (1990).

\bibitem{R94}
 B.~Ya. Ryabko,  The complexity and effectiveness of prediction
algorithms,
  \emph{J. Complexity}, \textbf{10}(3), 281--295, (1994).

\bibitem{RA}
 B.~ Ryabko, J.~ Astola and A.~ Gammerman,
  Application of Kolmogorov complexity and universal codes to identity testing and nonparametric testing of serial
  independence for time series,
  \textit{Theoretical Computer Science,} \textbf{359},  440-448, (2006).

\bibitem{RA2}
 B.~ Ryabko, J.~ Astola and A.~ Gammerman,
  Adaptive  Coding and Prediction  of Sources with Large and Infinite
  Alphabets, \textit{IEEE Transactions
on Information Theory}, \textbf{54}(8),   (2008).

\bibitem{RA3}
B. Ryabko, J.Astola and K. Egiazarian, Fast Codes for Large
Alphabets,
  \emph{Communications in Information and Systems},\textbf{ 3} (2),
  139--152, (2003).

  \bibitem{RM1}
 B.~ Ryabko and   V.~ Monarev,
  Experimental Investigation of Forecasting Methods Based on Data Compression Algorithms.
  \emph{Problems of Information Transmission,} \textbf{41}, (1), 65-69,
  (2005).

\bibitem{RM}
 B.~ Ryabko and   V.~ Monarev,  Using Information Theory Approach to  Randomness
 Testing,
\textit{Journal of Statistical Planning and Inference,}
\textbf{133}(1),  95--110, (2005).

\bibitem{RR}
 B.~ Ryabko and  Zh.~ Reznikova,  Using Shannon Entropy and Kolmogorov
Complexity
 To Study the Communicative System and Cognitive Capacities in
 Ants,
 \textit{Complexity,} \textbf{2} (2),  37--42, (1996).

 \bibitem{RT}
 B.~ Ryabko and   F.~ Topsoe,  On Asymptotically Optimal Methods of
Prediction and Adaptive Coding for Markov Sources, \textit{Journal
of Complexity,} \textbf{18}(1), 224--241, (2002).

\bibitem{DR}
 D.~ Ryabko and   M.~ Hutter, Sequence prediction for non-stationary
processes, In proceedings:  \textit{Combinatorial and Algorithmic
Foundations of Pattern and Association Discovery, } Dagstuhl
Seminar, 2006, Germany, http://www.dagstuhl.de/06201/  see also
http://arxiv.org/pdf/cs.LG/0606077

\bibitem{Sa}
  S.~A. Savari, A probabilistic approach to some
asymptotics in noiseless communication,  \textit{IEEE Transactions
on Information Theory}, \textbf{46}(4),  1246--1262,  (2000).


\bibitem{S}
C. ~E. Shannon, A mathematical theory of communication, \emph{
Bell Sys. Tech. J.} , {\bf 27},   379--423, 623--656, (1948).


\bibitem{S2}
C. ~E. Shannon, Communication theory of secrecy systems,
\emph{Bell Sys. Tech. J.}, {\bf 28},   656--715, (1948).

\bibitem{Shir}
A.N. Shiryaev, \emph{Probability}, (second edition), Springer,
1995.

\end{thebibliography}
\end{document}